\documentclass[a4paper,11pt]{article}
\pdfoutput=1
\usepackage{jheppub}[=2022--06-06]
\usepackage[T1]{fontenc}
\usepackage{lmodern}

\usepackage{microtype}

\newcommand{\U}{{\rm U}}

\newcommand*{\rdmathspace}[1][h]{%
\if h#1%
\thinmuskip=2mu\medmuskip=1mu\thickmuskip=3mu\fi%
\if m#1%
\thinmuskip=2.3mu\medmuskip=2mu\thickmuskip=3.5mu\fi%
\if s#1%
 \thinmuskip=3mu minus 0.7mu\medmuskip=4mu minus 2mu\thickmuskip=5mu plus 3.5mu minus 1.5mu\fi%
\if d#1%
\thinmuskip=3mu\medmuskip=4mu\thickmuskip=5mu plus 5mu\fi}

\usepackage{mathtools}
\usepackage{fancybox}
\usepackage{feynmp}

\usepackage{dcolumn}
\usepackage{stackengine}
\usepackage{dsfont}
\usepackage{enumitem}

\newcommand{\OO}{\mathcal{O}}
\newcommand{\lsim}{\lesssim}
\newcommand{\gsim}{\gtrsim}
\newcommand\branobar{\scalebox{.25}{\hspace{2pt}{\textbf (}}\scalebox{.25}{\hspace{17pt} {\textbf)}}}
\newcommand{\nunubar}{\stackon[-1.3pt]{$\bar\nu$}{\branobar}}
\newcommand{\SSbar}{\stackon[-1.2pt]{$\bar S\,$}{\branobar}}

\title{\boldmath Quasi-sterile neutrinos from dark sectors.
Part I. BSM matter effects in neutrino oscillations and the short-baseline anomalies.}

\author[a,1]{Daniele S.\;M. Alves,\note{Corresponding author.}}
\author[b]{William C. Louis}
\author[a]{and Patrick G. deNiverville}

\affiliation[a]{Theoretical Division, Los Alamos National Laboratory,\\Los Alamos, NM 87545, U.S.A.}
\affiliation[b]{Physics Division, Los Alamos National Laboratory,\\Los Alamos, NM 87545, U.S.A.}

\emailAdd{spier@lanl.gov}
\emailAdd{louis@lanl.gov}
\emailAdd{pgdeniverville@lanl.gov}

\abstract{Quasi-sterile neutrinos are a natural consequence of dark sectors interacting with the Standard Model (SM) sector via neutrino- and vector-portals.
Essentially, quasi-sterile neutrinos are light dark sector fermions with two generic properties: (i) they mix with the active neutrinos of the SM, and (ii) they are charged under a vector mediator that couples feebly to SM matter. Various interesting phenomenological consequences result from this class of particles. In this article, we investigate one such consequence: new, beyond the SM matter effects that can alter in-medium neutrino oscillations. In particular, for special windows of energy and matter densities, active neutrinos can resonantly oscillate into sterile neutrinos. We take advantage of this feature to build a quasi-sterile neutrino model that can explain the MiniBooNE and LSND anomalies, while remaining compatible with observations from long-baseline reactor- and accelerator-based neutrino experiments. This model is also likely compatible with the recent results reported by the MicroBooNE collaboration (albeit we cannot precisely quantify this claim due to a lack of information in MicroBooNE's public data releases to date). Implications for solar neutrinos and $\nu_e$ disappearance searches are also briefly discussed.}
\providecommand\inspire[1]{\href{https://inspirehep.net/search?p=find+#1}{{\tiny IN}{\footnotesize SPIRE}}}

\providecommand\erratum[4][ibid.\ ]{\emph{Erratum #1}{\bf #2} (#3) #4}

\providecommand{\jhep}[3] {\ifnum#2>2009%
\href{https://doi.org/10.1007/JHEP#1(#2)#3}{\emph{JHEP} {\bf #1} (#2) #3}%
\else%
\href{https://doi.org/10.1088/1126-6708/#2/#1/#3}{\emph{JHEP} {\bf #1} (#2) #3}%
\fi}
\providecommand{\jcap}[3] {\href{https://doi.org/10.1088/1475-7516/#2/#1/#3}{\emph{JCAP} {\bf #1} (#2) #3}} 
\def\issueFromCounter.#1#2#3#4#5#6.{#2#3}
\providecommand{\jstat}[2]{\PackageWarningNoLine{\jname}{The macro \protect\jstat\space is obsolete!\MessageBreak Please typeset JSTAT as any other journal}%
  \href{https://doi.org/10.1088/1742-5468/#1/\issueFromCounter.#2./#2}{\emph{J.\ Stat.\ Mech.\ }(#1) #2}} 

\providecommand{\hepph}[1]{\href{https://arxiv.org/abs/hep-ph/#1}{\tt hep-ph/#1}}

\providecommand{\hepex}[1]{\href{https://arxiv.org/abs/hep-ex/#1}{\tt hep-ex/#1}}

\providecommand{\nuclex}[1]{\href{https://arxiv.org/abs/nucl-ex/#1}{\tt nucl-ex/#1}}

\providecommand{\arXivid}[1]{\href{https://arxiv.org/abs/#1}{\tt arXiv:#1}}
\providecommand{\Math}[2]{%
\if!#1!%
\href{https://arxiv.org/abs/math/#2}{\tt math/#2}%
\else%
\href{https://arxiv.org/abs/math.#1/#2}{\tt math.#1/#2}%
\fi}

\begin{document}
\preprint{LA-UR-22-20015}
\maketitle
\flushbottom

\section{Introduction}
\label{intro}

\begin{figure}
\centering
\includegraphics[width=0.7\textwidth]{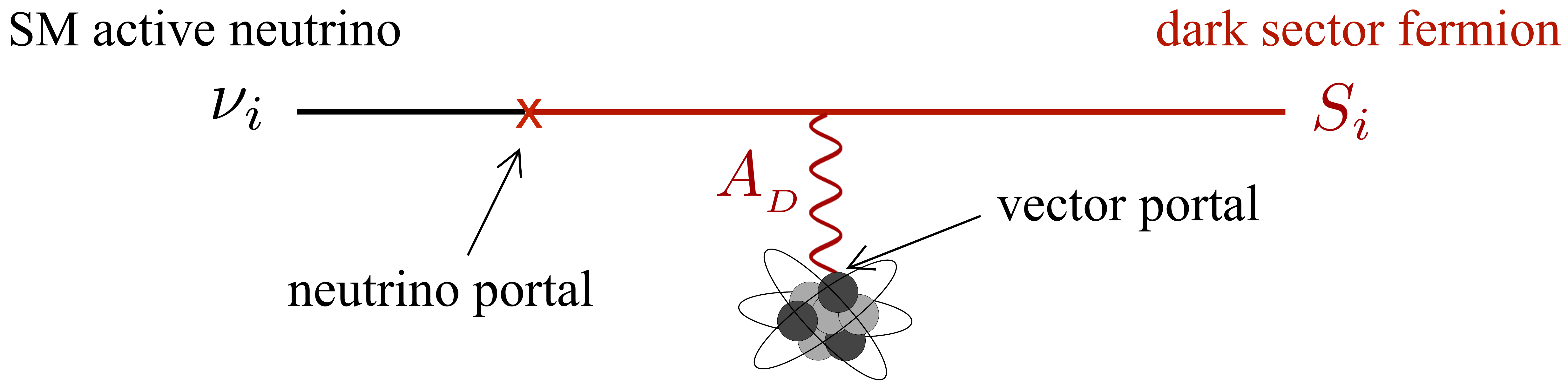}
\caption{A generic consequence of dark sectors with neutrino- and vector-portals are quasi-sterile neutrinos. The BSM matter potential generated by the dark vector interactions can alter neutrino oscillations in~matter.\label{Cartoon_QuasiSterile}}
\end{figure}

Over the past decade, light dark sectors have emerged as a contender to the WIMP paradigm in providing a possible explanation for the cosmological origin and particle nature of dark matter~\cite{Arkani-Hamed-ml-2008hhe,Jaeckel-ml-2010ni,Essig-ml-2013lka,Alexander-ml-2016aln,Lanfranchi-ml-2020crw}. While there is no consensus on a precise definition of dark sectors, a fairly generic framework to characterize their phenomenology are the so-called \emph{portals}: particle `mediators' that couple to matter and/or forces both in the Standard Model (SM) and in the dark sector, typically (but not necessarily) by mixing with neutral SM particles --- such as the photon, the Higgs boson, or neutrinos --- if their quantum numbers permit. Most dark sector phenomenological studies, however, take the approach of `switching on' only one individual portal at a time. In this article, we investigate a broader scenario of two simultaneous dark sector portals, namely, the \emph{neutrino} and \emph{vector} portals, and focus on one generic phenomenological consequence: \emph{quasi-sterile neutrinos}. Specifically, the neutrino portal can be realized by light (say, sub-keV) dark sector fermions mixing with the active neutrinos of the Standard Model. Since these dark sector fermions are neutral under the SM gauge interactions, they are, by definition, \emph{sterile neutrinos}. However, in the presence of a vector portal --- a dark sector gauge interaction that couples feebly to SM fermions --- sterile neutrinos charged under the vector portal might be more appropriately characterized as \emph{quasi}-sterile neutrinos (see figure~\ref{Cartoon_QuasiSterile}). From this, two phenomenological features naturally arise: (i) the presence of beyond the Standard Model (BSM) neutrino oscillations due to mixing of active and sterile neutrinos, and (ii) the possibility of BSM matter effects, and therefore, of \emph{resonant} active-to-sterile neutrino oscillations for specific regimes of matter density and neutrino energy.

There are many interesting motivations to study quasi-sterile neutrinos. One of them is the excess of electron recoil events recently reported by the XENON1T collaboration~\cite{XENON-ml-2020rca}. Attempts to explain this excess as a signal of solar neutrinos scattering off electrons via an anomalous neutrino magnetic dipole moment are in tension with limits from stellar cooling~\cite{Corsico-ml-2014mpa,bib2019arXiv191010568A,Yusof-ml-2021fhu,Studenikin-ml-2021fmn}. On the other hand, these astrophysical constraints could be relaxed if the anomalous XENON1T electron recoil events were instead induced by quasi-sterile solar neutrinos scattering via a light dark mediator (see also~\cite{Boehm-ml-2020ltd,Bloch-ml-2020uzh,Babu-ml-2020ivd,Ge-ml-2021snv}). This topic will be explored in a future publication~\cite{future}.

Another motivation for quasi-sterile neutrinos, which will be the focus of this article, is the puzzle posed by the short-baseline neutrino oscillation anomalies, which
could be an indication of light sterile states mixing with the active neutrinos of the SM\@.
Two intriguing anomalies of this type are the $\bar{\nu}_\mu\to\bar{\nu}_e$ and $\nu_\mu\to \nu_e$  appearance events reported by the LSND~\cite{LSND-ml-1996ubh,LSND-ml-1997vun,LSND-ml-1996vlr,LSND-ml-1997vqj,LSND-ml-2001aii} and MiniBooNE~\cite{MiniBooNE-ml-2018esg,MiniBooNE-ml-2020pnu} collaborations. The data in both of these experiments show excesses of (anti-)electron-neutrino charged-current scattering events at $L/E_\nu\sim 0.5-2$ m/MeV, at rates compatible with
$\nunubar_{\!e}$ appearance probabilities of \emph{O}(0.1--1\%). These $\nunubar_{\!e}$ appearance events cannot be accommodated in the SM description of neutrino masses and mixings. A simple BSM model including one sterile neutrino species in addition to the three active species (the `vanilla' 3+1 model) can provide a good fit to the LSND excess, the MiniBooNE excess in \emph{antineutrino} mode, and the MiniBooNE excess in \emph{neutrino} mode \emph{for the energy range} $E_\nu\gsim400$\,MeV\@. However, an excess of $\nu_e$ appearance events with lower energies $E_\nu\sim150-300$\,MeV still persists in the MiniBooNE data which cannot be satisfactorily explained by any `3+$n$' sterile neutrino model\footnote{Recent results from the MicroBooNE collaboration~\cite{MicroBooNE-ml-2021zai} further disfavor the interpretation of the MiniBooNE low energy excess as due to single photon backgrounds from radiative $\Delta$ decays~\cite{Hill-ml-2010zy,Zhang-ml-2012xn,Wang-ml-2013wva,Wang-ml-2014nat,Alvarez-Ruso-ml-2021dna}.}~\cite{Maltoni-ml-2002xd,Maltoni-ml-2007zf,Conrad-ml-2012qt,Kopp-ml-2013vaa,Giunti-ml-2015mwa,Collin-ml-2016rao,Boser-ml-2019rta,Alvarez-Ruso-ml-2021dna,Brdar-ml-2021cgb}. This challenge has led to numerous attempts to explain the MiniBooNE low energy excess by invoking new physics of a dark sector. In the `non-oscillatory' approaches~\cite{Gninenko-ml-2009ks,Masip-ml-2012ke,Bai-ml-2015ztj,Jordan-ml-2018qiy,Ballett-ml-2018ynz,Bertuzzo-ml-2018itn,Arguelles-ml-2018mtc,Ballett-ml-2019pyw,Fischer-ml-2019fbw,deGouvea-ml-2019qre,Hostert-ml-2020oui,Datta-ml-2020auq,Abdallah-ml-2020biq,Abdullahi-ml-2020nyr,Dutta-ml-2021cip,Vergani-ml-2021tgc}, production and decays of dark sector fermions and/or light bosons yield final states with $\nu_e$, $\gamma$, or collimated $e^+e^-$ pairs in the MiniBooNE detector, all of which could potentially mimic a $\nu_e$ charged-current scattering event. In the `oscillatory' approaches, BSM effects alter neutrino propagation through a medium (either ordinary matter~\cite{Karagiorgi-ml-2012kw}, the C$\nu$B~\cite{Asaadi-ml-2017bhx}, or extra dimensions~\cite{Pas-ml-2005rb,Doring-ml-2018cob}), leading to resonant neutrino oscillations within the window of energy where the MiniBooNE low energy excess is concentrated. Previous attempts to implement oscillatory approaches have not been successful, however, because new matter effects invariably alter the effective in-medium mass splittings of active neutrinos~\cite{Kopp-ml-2014fha,Denton-ml-2018dqq,Barenboim-ml-2019hso,Smirnov-ml-2021zgn}, leading to oscillation probabilities incompatible with observations at long-baseline neutrino experiments.

In this article, we show that these difficulties can be overcome by a `triple-resonance mechanism' that is able to preserve the in-medium active neutrino mass spectrum at high energies probed by accelerator-based neutrino experiments, as well as the in-medium antineutrino mass spectrum at low energies probed by reactor-based experiments.

At this point it is incumbent on us to comment on the recent $\nu_e$ searches by the MicroBooNE experiment based on data obtained from an exposure of $7\times 10^{20}$ protons-on-target (POT) from the Booster Neutrino Beam (BNB)~\cite{MicroBooNE-ml-2021rmx}. These analyses appear to strongly disfavor the hypothesis that the MiniBooNE low energy excess is due to $\nu_e$ scattering events. As we shall discuss in subsection~\ref{MicroBooNE}, this conclusion is strongly model dependent and likely does not hold at a high enough confidence level to exclude the `oscillatory' BSM explanation of the MiniBooNE low energy excess proposed in this study.

This article is organized as follows: in section~\ref{toymodel}, we introduce and dissect the main features of the resonant mechanism in a toy model, which will then be used in section~\ref{FullModel} to build a full model explaining the MiniBooNE excess. Section~\ref{SunMicroBooNE} considers the full model's consistency with the recent MicroBooNE results, and provides a brief discussion of implications for solar neutrinos. Section~\ref{Conclusion} summarizes this article's main points and provides an outlook of future experimental measurements that could falsify the quasi-sterile neutrino explanation of short-baseline anomalies.

\section{The minimal case as a toy model}
\label{toymodel}

We begin by considering the simplest case of a single quasi-sterile neutrino flavor, which we shall call $S_3$, mixing with the active neutrinos of the SM model. It turns out that this minimalistic model does not provide a satisfactory explanation of the short-baseline neutrino anomalies. However, it retains all the basic phenomenological features of the full model to be considered in the next section, and therefore, for pedagogical purposes, we will introduce it first in this section and highlight its key aspects and shortcomings.

We adopt the PDG convention~\cite{ParticleDataGroup-ml-2020ssz} to relate the neutrino mass eigenstates in the SM, $\nu_i$ ($i=1,2,3$), to the active neutrino flavors, $\nu_\ell$ ($\ell=e,\mu,\tau$):
\begin{subequations}\label{SMeigenstates}
\begin{alignat}{3}
&|\nu_1\rangle~&=~&U_{e1}\,|\nu_e\rangle~+~U_{\mu 1}\,|\nu_\mu\rangle~+~U_{\tau 1}\,|\nu_\tau\rangle \\
&|\nu_2\rangle~&=~&U_{e2}\,|\nu_e\rangle~+~U_{\mu 2}\,|\nu_\mu\rangle~+~U_{\tau 2}\,|\nu_\tau\rangle \\
&|\nu_3\rangle~&=~&U_{e3}\,|\nu_e\rangle~+~U_{\mu 3}\,|\nu_\mu\rangle~+~U_{\tau 3}\,|\nu_\tau\rangle
\end{alignat}
\end{subequations}
where the $U_{\ell i}$ coefficients above are the matrix elements of the PMNS matrix $U$, which is parametrized by three mixing angles $\theta_{12}$, $\theta_{23}$ and $\theta_{13}$, and a CP-violating phase $\delta_{\text{CP}}$:
\begin{equation}
U = \begin{pmatrix}
1 & ~0 & ~0 \\
0 & ~c_{23} & ~s_{23} \\
0 &-s_{23} & ~c_{23}
\end{pmatrix}\cdot
\begin{pmatrix}
~c_{13} & ~0 & ~~s_{13}\,e^{-i\delta_\text{CP}} \\
0 & ~1 & ~0 \\
-s_{13}\,e^{i\delta_\text{CP}} & ~0 & ~c_{13}
\end{pmatrix}\cdot
\begin{pmatrix}
~c_{12} & s_{12} & ~0 \\
-s_{12} & c_{12} & ~0 \\
~0 & 0 & ~1
\end{pmatrix}
\end{equation}
with $s_{ij}\equiv\sin\theta_{ij}$ and $c_{ij}\equiv\cos\theta_{ij}$.

In this minimal model, we will introduce a small mixing between the sterile neutrino $S_3$ and one of the active neutrino states, namely, $\nu_3$, parametrized by a mixing angle $\theta_{_{\!S_3}}\ll 1$.
The mass eigenstates \emph{in vacuum} are given by:
\begin{subequations}\label{vacuumEigenstates}
\begin{alignat}{4}
&|\nu_1\rangle~&=&~\sum_{~~~\ell\,=\,e,\,\mu,\,\tau} \!\!\!U_{\ell 1}\,|\nu_\ell\rangle\\
&|\nu_2\rangle~&=&~\sum_{~~~\ell\,=\,e,\,\mu,\,\tau} \!\!\!U_{\ell 2}\,|\nu_\ell\rangle\\
&|N_{3,\text{ vac}}^{_{(-)}}\rangle~&=&-e^{i\,\delta_{_{\!S_3}}}\sin\theta_{_{\!S_3}}\,|S_3\rangle \;+\; \cos\theta_{_{\!S_3}}\,|\nu_3\rangle\\
&&=&-e^{i\,\delta_{_{\!S_3}}}\sin\theta_{_{\!S_3}}\,|S_3\rangle \;+\; \cos\theta_{_{\!S_3}}\!\!\!\!\!\sum_{~~~\ell\,=\,e,\,\mu,\,\tau}\!\!\!\!U_{\ell 3}\,|\nu_\ell\rangle \nonumber\\
&|N_{3,\text{ vac}}^{_{(+)}}\rangle~&=&~\cos\theta_{_{\!S_3}}\,|S_3\rangle \;+\; e^{-i\,\delta_{_{\!S_3}}}\sin\theta_{_{\!S_3}}\,|\nu_3\rangle\\
&&=&~\cos\theta_{_{\!S_3}}\,|S_3\rangle \;+\; e^{-i\,\delta_{_{\!S_3}}}\sin\theta_{_{\!S_3}}\!\!\!\!\!\sum_{~~~\ell\,=\,e,\,\mu,\,\tau}\!\!\!\!U_{\ell 3}\,|\nu_\ell\rangle \nonumber
\end{alignat}
\end{subequations}
with mass eigenvalues $m_1$, $m_2$, $m_3$, and $M_3\gg m_3$, respectively. Since $\sin\theta_{_{\!S_3}}\ll 1$, the heavy eigenstate $N_3^{_{(+)}}$ is mostly sterile with a small active component, whereas $N_3^{_{(-)}}$ is composed mostly of the active neutrino state $\nu_3$, with a small sterile component. In addition, note that because $S_3$ mixes exclusively with $\nu_3$, the complex phase $\delta_{_{\!S_3}}$ is unphysical and can be fully reabsorbed by a unitary rotation $|S_3\rangle\to e^{-i\,\delta_{_{\!S_3}}} |S_3\rangle$. We will return to the issue of additional CP-violating phases in subsection~\ref{VanillaPart}.

 All of the above information can be captured in the neutrino Hamiltonian. Working in the basis $B=\{|\nu_1\rangle,|\nu_2\rangle,|\nu_3\rangle,|S_3\rangle\}$, we can write the vacuum Hamiltonian in a familiar~form:
{\rdmathspace[h]
\begin{align}\label{vacuumH}
\widehat{H}_{B,\text{ vacuum}} &\simeq
\begin{pmatrix}
1 & ~~0~ & 0 & ~0 \\
0 & ~~1~ & 0 & ~0 \\
0 & ~~0~ & ~~\cos\theta_{_{\!S_3}} & \sin\theta_{_{\!S_3}} \\
0 & ~~0 & - \sin\theta_{_{\!S_3}} & \cos\theta_{_{\!S_3}}
\end{pmatrix}\cdot
\begin{pmatrix}
\frac{m_1^2}{2\,E_\nu} & ~0 & ~0 & ~0 \\
0~ &\frac{m_2^2}{2\,E_\nu} & ~0 & ~0 \\
0~ & 0~ & \frac{m_3^2}{2\,E_\nu} & ~0 \\
0~ & 0~ & 0~ & \frac{M_3^2}{2\,E_\nu}
\end{pmatrix}\cdot
\begin{pmatrix}
1 & ~~0~~ & 0 & ~0 \\
0 & ~~1~~ & 0 & ~0 \\
0 & ~~0~~ & \cos\theta_{_{\!S_3}} & -\sin\theta_{_{\!S_3}} \\
0 & ~~0~~ & \sin\theta_{_{\!S_3}} & ~~\cos\theta_{_{\!S_3}}
\end{pmatrix}
\nonumber\\ &= (\ldots )\times \mathds{1}_{3\times 3}~+~ \frac{\delta M_3^2}{4E_\nu}
\begin{pmatrix}
-1+2 \frac{\delta m_{13}^2}{\delta M_3^2} & 0 & 0 & 0 \\
0 & -1+ 2\frac{\delta m_{23}^2}{\delta M_3^2} & 0 & 0 \\
0 & 0 & -\cos2\theta_{_{\!S_3}}~ & \sin2\theta_{_{\!S_3}} \\
0 & 0 & ~~\sin2\theta_{_{\!S_3}}~ & \cos2\theta_{_{\!S_3}}
\end{pmatrix}
\end{align}
}\relax where we have defined the mass-squared splittings $\delta M_3^2 \equiv M_3^2-m_3^2$, ~$\delta m_{ij}^2 \equiv m_i^2-m_j^2$, and we did not bother to write the terms proportional to the identity operator in the second line of~(\ref{vacuumH}), since they do not affect the difference between energy eigenvalues and therefore do not contribute to neutrino oscillations.

When propagating through a medium, neutrinos receive additional contributions to their energies due to the presence of matter potentials (which can be understood as coherent forward scattering at zero momentum transfer). If this effect is felt identically by all neutrino states --- which is the case, in the SM, of the matter potential generated by the neutral weak interactions mediated by the $Z^0$ boson --- the relative phases of the neutrino propagation eigenstates are not changed, and therefore such matter potentials do not affect neutrino oscillations in a medium. On the other hand, a matter potential that affects different neutrino flavors non-identically will alter the propagation eigenstates and their relative phases with respect to vacuum, and can lead to significant distortions in neutrino oscillations for certain energy and medium density regimes. The most well-known such effect is due to the charged weak interactions mediated by the $W^\pm$ bosons, which generate an effective matter potential to the electron-neutrino flavor, but not to muon- nor tau-neutrino flavors, since the only charged leptons present in ordinary matter are electrons. This is known as the standard Mikheyev–Smirnov–Wolfenstein (MSW) effect~\cite{PhysRevD.17.2369,Mikheyev-ml-1985zog,PhysRevD.22.2718,Smirnov-ml-2004zv}.

In the framework of dark sectors, it is only natural to expect that dark gauge bosons might mediate interactions between sterile neutrinos and SM fermions. The interactions with stable SM fermions, specifically, will generate a new matter potential source for neutrinos propagating through matter, and will alter neutrino oscillations as long as the dark gauge boson does not couple universally to all active and sterile neutrino species. As an example, consider such a dark vector mediator $A_{\!_D}$ coupling to the following currents:
\begin{align}\label{Lmediator}
\mathcal{L}_\text{dark} \supset - A_{{\!_D}}^\mu&\Bigg(g_{_S}\,\bar{S}_3\;\frac{\!\gamma_\mu(1-\gamma_5)\!\!}{2}\;S_3 + \sum_{i}g_\nu \;\bar{\nu}_i\;\frac{\!\gamma_\mu(1-\gamma_5)\!\!}{2}\;\nu_i \nonumber\\
&\; + g_e\,\bar{e}\,\gamma_\mu \,e + g_p \,\bar{p}\,\gamma_\mu \,p + g_n \,\bar{n}\,\gamma_\mu\, n + \cdots\Bigg)\,,
\end{align}
where the dots denote unspecified mediator couplings to heavier fermions. The resulting matter potential experienced by (anti-)neutrinos propagating through matter is then given by (see, e.g.,~\cite{Linder-ml-2005fc}):
\begin{subequations}\label{Vm0}
\begin{alignat}{3}
&V_{\nu_i}\big|_\text{matter} &=  - V_{\overline\nu_i}\big|_\text{matter} &=  - \frac{\; g_\nu }{\;2\,m_{A_{\!_D}}^2\!\!}\;\!\!\!\!\!\sum_{~~~f = e,\,p,\,n}\!\!\!\!g_f\,n_f\,, \\[5pt]
&V_{S_3}\big|_\text{matter} &=  - V_{\overline{S}_3}\big|_\text{matter} &=  - \frac{\; g_{_S} }{\;2\,m_{A_{\!_D}}^2\!\!}\;\!\!\!\!\!\sum_{~~~f = e,\,p,\,n}\!\!\!\!g_f\,n_f\,.
\end{alignat}
\end{subequations}
Above, $m_{A_{\!_D}}$ is the dark mediator's mass, and $n_f$ is the in-medium number density of the SM fermion species $f$. It is clear from~(\ref{Vm0}) that these $A_{\!_D}$-induced matter potentials are non-zero only if ordinary matter has finite `dark charge' density. The latter requirement rules out the dark photon as a mediator candidate, since ordinary matter is, on average, charge-neutral to a dark photon. In this specific example, the three active neutrinos experience the same matter potential $V_{\nu_i}$, and therefore SM neutrino oscillations are unaffected in regimes for which BSM effects can be neglected. However, as long as $g_{_S}\neq g_\nu$, the matter potentials for the sterile and active species will be distinct, and therefore neutrino oscillations will be affected by the relative potential experienced by $S_3$:
\begin{equation}\label{Vm}
\Delta V\big|_\text{matter} = (V_{S_3} - V_{\nu_i})\big|_\text{matter} =  - \frac{\; (g_{_S} - g_\nu) }{\;2\,m_{A_{\!_D}}^2\!\!}\;\!\!\!\!\!\sum_{~~~f = e,\,p,\,n}\!\!\!\!g_f\,n_f\,.
\end{equation}

We note that a variety of phenomenological constraints restrict the mass and couplings of the dark mediator. For instance, $S_3$ in-medium scattering with finite momentum exchange must be sufficiently suppressed, otherwise this would lead to decoherence of $S_3-\nu_3$ oscillations~\cite{Nieves-ml-2020jjg}. Further, the combinations $g_p(g_\nu-g_{_S}\sin\theta_{_{\!S_3}})$, $g_n(g_\nu-g_{_S}\sin\theta_{_{\!S_3}})$, and $g_e(g_\nu-g_{_S}\sin\theta_{_{\!S_3}})$ are constrained by neutrino-proton, neutrino-neutron, and neutrino-electron scattering, respectively~\cite{Kolb-ml-1987qy,Bilmis-ml-2015lja,COHERENT-ml-2017ipa,Coloma-ml-2020gfv}. These constraints can be respected for sufficiently small couplings, and, as evident from~(\ref{Vm}), it is possible to simultaneously keep $\Delta V$ non-negligible if the mediator mass $m_{A_{\!_D}}$ is sufficiently light. Numerous other constraints from colliders, beam dumps, fixed targets, meson decays, stellar cooling, anomalous magnetic dipole moments, etc., also apply in different regions of parameter space of~(\ref{Lmediator})~\cite{Ilten-ml-2018crw,Fabbrichesi-ml-2020wbt,Caputo-ml-2021eaa,Agrawal-ml-2021dbo}. We will, however, defer these considerations to a future work. For the remainder of this article, we will remain agnostic about the mass and couplings of the dark mediator $A_{\!_D}$, and instead treat the effective matter potential $\Delta V$ as a free parameter in order to address the MiniBooNE anomaly. A comprehensive study of experimentally viable UV completions of $V_{\nu_i}$, $V_{S_3}$, and further signal predictions, will be the focus of an upcoming publication~\cite{future}.

The effective matter potential~(\ref{Vm}) for $S_3$ modifies the neutrino Hamiltonian in matter:
\begin{equation}\label{Hmatter}
\widehat{H}_{B,\text{ matter}} = \widehat{H}_{B,\text{ vacuum}} -
\begin{pmatrix}
~0 & ~~~0 & ~~~0 & \,0 \\
~0 & ~~~0 & ~~~0 & \,0 \\
~0 & ~~~0 & ~~~0 & \,0 \\
~0 & ~~~0 & ~~~0 & |\Delta V| \!
\end{pmatrix}\,,
\end{equation}
where we chose the sign of $\Delta V$ to be negative (positive) for neutrinos (antineutrinos) for reasons that will become clear shortly.

The neutrino propagation eigenstates of $\widehat{H}_{B,\text{matter}}$ in~(\ref{Hmatter}) and their eigenvalues $\omega$ are easily obtained:
\begin{subequations}\label{matterEigen}
\begin{alignat}{3}
&\!\!\!\!\!\! |N_{1,\,m}\rangle~=~|\nu_1\rangle\,, &\omega_1~~&=~\frac{\delta m^2_{13}}{2 E_\nu} \label{matterEigen1}\\
\nonumber\\
&\!\!\!\!\!\! |N_{2,\,m}\rangle~=~|\nu_2\rangle\,, &\omega_2~~&=~\frac{\delta m^2_{23}}{2 E_\nu} \label{matterEigen2}\\
 \nonumber\\
&\!\!\!\!\!\! |N_{3,\, m}^{_{(-)}}\rangle~=-\sin\theta_{_{\!S_3}}^{_{(m)}}\,|S_3\rangle \,+\, \cos\theta_{_{\!S_3}}^{_{(m)}}\,|\nu_3\rangle\,,\qquad &\omega_3^{_{(-)}} &=~\frac{\delta M_3^2}{4E_\nu}(1-C_3)-\frac{|\Delta V|}{2} \label{matterEigen3}\\
\nonumber\\
&\!\!\!\!\!\!  |N_{3,\, m}^{_{(+)}}\rangle~=~~\cos\theta_{_{\!S_3}}^{_{(m)}}\,|S_3\rangle \,+\, \sin\theta_{_{\!S_3}}^{_{(m)}}\,|\nu_3\rangle\,,\qquad &\omega_3^{_{(+)}} &=~\frac{\delta M_3^2}{4E_\nu}(1+C_3)-\frac{|\Delta V|}{2} \label{matterEigen4}\\
\nonumber
\end{alignat}
\end{subequations}
where $C_3$ and the effective $\nu_3-S_3$ mixing angle in matter, $\theta_{_{\!S_3}}^{_{(m)}}$, are defined through:
\begin{equation}\label{effMixAngle}
\sin2\theta_{_{\!S_3}}^{_{(m)}} = \frac{\sin2\theta_{_{\!S_3}}}{C_3} \equiv \frac{\sin2\theta_{_{\!S_3}}}{\;\sqrt{~\Big(\!\sin2\theta_{_{\!S_3}\!}\Big)^2 + \left(\cos2\theta_{_{\!S_3}} - \frac{4E_\nu}{\delta M_3^2}\frac{|\Delta V|}{2}\right)^2~}\;}\,.
\end{equation}

Expressions analogous to~(\ref{matterEigen1})--(\ref{matterEigen4}) and~(\ref{effMixAngle}) hold for antineutrinos, with one important difference: $-|\Delta V|$ should be replaced everywhere by $+|\Delta V|$.
Note, importantly, that because of our choice of sign for the matter potential $\Delta V$, $\nu_3- S_3$ oscillations become \emph{resonant} (i.e., the mixing angle becomes maximal, $\theta_{_{\!S_3}}^{_{(m)}}=\pi/4$) when $E_\nu$ is at the resonant energy value of
\begin{equation}
\label{Eres3}
E^\text{res}_{\nu_3} = \frac{\delta M_3^2\,\cos2\theta_{_{\!S_3}}}{2\,|\Delta V|}\,.
\end{equation}
Because the matter potential for antineutrinos has the opposite sign, no such resonant effect is present for $\overline\nu_3-\overline{S}_3$ oscillations.

\begin{figure}
\centering
\includegraphics[width=0.7\textwidth,trim= 0 0 20 0,clip]{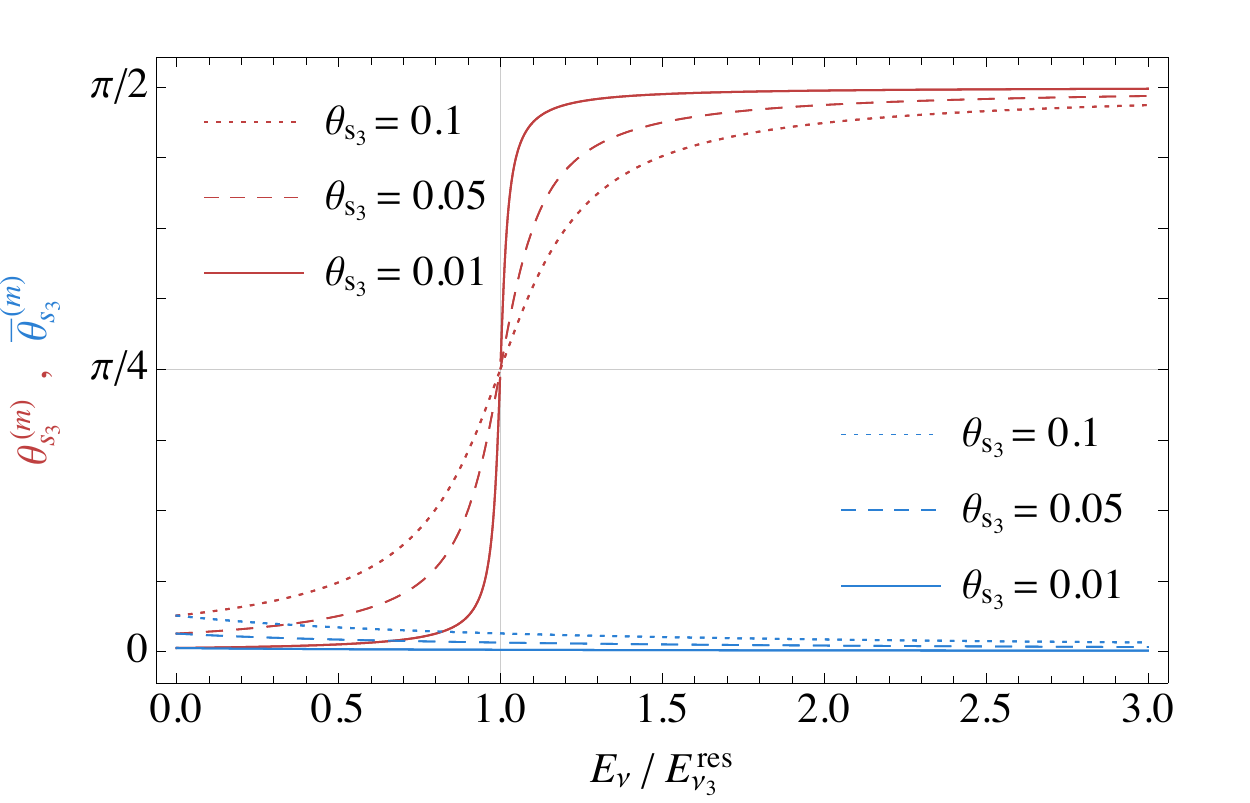}
\caption{The in-medium effective mixing angles $\theta_{_{\!S_3}}^{_{(m)}}$ (red curves) and $\bar\theta_{_{\!S_3}}^{_{(m)}}$ (blue curves) between $\nu_3-S_3$ and $\bar\nu_3-\bar S_3$, respectively. The dependence of these angles is shown as a function of (anti)neutrino energy $E_\nu$ in units of the resonant energy $E^\text{res}_{\nu_3}$ for different choices of the vacuum mixing angle $\theta_{_{\!S_3}}$.\label{Theta3m}}
\end{figure}

Figure~\ref{Theta3m} displays the effective active-sterile mixing angles in matter for both neutrinos ($\theta_{_{\!S_3}}^{_{(m)}}$) and antineutrinos ($\bar\theta_{_{\!S_3}}^{_{(m)}}$) as a function of (anti-)neutrino energy, under three illustrative assumptions for the vacuum mixing angle $\theta_{_{\!S_3}}$. Note that the transition from $\theta_{_{\!S_3}}^{_{(m)}}\approx\theta_{_{\!S_3}}$ to $\theta_{_{\!S_3}}^{_{(m)}}\to\pi/2$ as $E_\nu$ crosses the resonant region becomes sharper for smaller values of the vacuum mixing angle $\theta_{_{\!S_3}}$. For antineutrinos, on the other hand, the active-sterile mixing angle $\bar\theta_{_{\!S_3}}^{_{(m)}}$ decreases monotonically with increasing energy (see also subsection~\ref{ToyAntiNU}).

\subsection{Neutrino propagation eigenstates in three energy regimes}
\label{regimes}
\enlargethispage{\baselineskip}

With~(\ref{matterEigen1})--(\ref{matterEigen4}) and~(\ref{effMixAngle}), we can now proceed to discuss the composition of the propagation eigenstates at finite matter density, and their respective oscillation phases, in three energy regimes relative to the resonant energy $E^\text{res}_{\nu_3}$ defined in~(\ref{Eres3}).
\begin{itemize}
\item
{\bf At very low energies}, $E_\nu\ll E^\text{res}_{\nu_3}$, or alternatively, at very low medium densities, $|\Delta V| \to0 $, we recover the vacuum oscillation results, with the mixing angle between $\nu_3$ and $S_3$ being
\begin{equation}
\theta_{_{\!S_3}}^{_{(m)}} \simeq  \theta_{_{\!S_3}}
\end{equation}
and the propagation eigenstates and respective oscillation phases given by:
\begin{subequations}
\begin{alignat}{3}
&\!\!\!\!\!\! |N_{1,\,m}\rangle~=~|\nu_1\rangle\,, &\omega_1~~&=\,~~\frac{\delta m^2_{13}}{2 E_\nu} \\
&\!\!\!\!\!\! |N_{2,\,m}\rangle~=~|\nu_2\rangle\,, &\omega_2~~&=\,~~\frac{\delta m^2_{23}}{2 E_\nu} \\
&\!\!\!\!\!\! |N_{3,\, m}^{_{(-)}}\rangle~=-\sin\theta_{_{\!S_3}}\,|S_3\rangle \,+\, \cos\theta_{_{\!S_3}}\,|\nu_3\rangle\,,~~ &\omega_3^{_{(-)}}&\simeq-\,\frac{\,\sin^2\!\theta_{_{\!S_3}}\delta M_3^2\,}{2E_\nu}\!\left(\!\frac{E_\nu}{\,E^\text{res}_{\nu_3}\,}\!\cos 2\theta_{_{\!S_3}}\!\right) \label{omega3MinuslowE}  \\
&\!\!\!\!\!\!  |N_{3,\, m}^{_{(+)}}\rangle~=~~\cos\theta_{_{\!S_3}}\,|S_3\rangle \,+\, \sin\theta_{_{\!S_3}}\,|\nu_3\rangle\,,~~&\omega_3^{_{(+)}}&\simeq\,~~\frac{\delta M_3^2}{\,2E_\nu\,}\!\left(\!1-\frac{E_\nu}{\,E^\text{res}_{\nu_3}\,}\!\cos 2\theta_{_{\!S_3}}\!\right)
\end{alignat}
\end{subequations}

\item
{\bf At the resonance}, $E_\nu = E^\text{res}_{\nu_3}$, the mixing between $\nu_3$ and $S_3$ is maximal,
\begin{equation}
\theta_{_{\!S_3}}^{_{(m)}}  =  \frac{\pi}{4}\,,
\end{equation}
and the propagation eigenstates and respective oscillation phases are given by:
 \begin{subequations}
\begin{alignat}{3}
&\!\!\!\!\!\!\!\! |N_{1,\,m}\rangle~=~|\nu_1\rangle\,, &\omega_1~~&=\,~~\frac{\delta m^2_{13}}{2 E_\nu} \\
&\!\!\!\!\!\!\!\! |N_{2,\,m}\rangle~=~|\nu_2\rangle\,, &\omega_2~~&=\,~~\frac{\delta m^2_{23}}{2 E_\nu} \\
&\!\!\!\!\!\!\!\! |N_{3,\, m}^{_{(-)}}\rangle\,=~\frac{1}{\sqrt{2}}\,\big(\,|\nu_3\rangle \,-\, |S_3\rangle\,\big)\,,\qquad\qquad\qquad &\omega_3^{_{(-)}}&\simeq\, -\,\frac{\,\sin \theta_{_{\!S_3}}\,\delta M_3^2\,}{2E_\nu} \qquad  \\
&\!\!\!\!\!\!\!\!  |N_{3,\, m}^{_{(+)}}\rangle\,=~\frac{1}{\sqrt{2}}\,\big(\,|\nu_3\rangle \,+\, |S_3\rangle\,\big)\,,\qquad\qquad\qquad &\omega_3^{_{(+)}}&\simeq\, +\,\frac{\,\sin \theta_{_{\!S_3}}\,\delta M_3^2\,}{2E_\nu}
\end{alignat}
\end{subequations}

\item
{\bf At very high energies}, $E_\nu\gg E^\text{res}_{\nu_3}$, the mixing between $\nu_3$ and $S_3$ is strongly suppressed, with
\begin{equation}
\theta_{_{\!S_3}}^{_{(m)}}  \to  \frac{\pi}{2}\,,
\end{equation}
so that their hierarchy is inverted, the active neutrino $\nu_3$ becomes the heavier eigenstate, and the sterile neutrino $S_3$ becomes the lighter eigenstate:
 \begin{subequations}
\begin{alignat}{3}
&\!\!\!\!\!\! |N_{1,\,m}\rangle~~=\,~~~|\nu_1\rangle\,, &\omega_1~~&=~\frac{\delta m^2_{13}}{2 E_\nu} \\
&\!\!\!\!\!\! |N_{2,\,m}\rangle~~=\,~~~|\nu_2\rangle\,, &\omega_2~~&=~\frac{\delta m^2_{23}}{2 E_\nu} \\
&\!\!\!\!\!\!  |N_{3,\, m}^{_{(-)}}\rangle\,~\to -\,|S_3\rangle\,,\qquad\qquad\qquad &\omega_3^{_{(-)}}&\simeq~\frac{\delta M_3^2}{2E_\nu}\!\left(\!1-\frac{E_\nu}{\,E^\text{res}_{\nu_3}\,}\!\cos 2\theta_{_{\!S_3}}\!\right) \\
&\!\!\!\!\!\! |N_{3,\, m}^{_{(+)}}\rangle\,~\to\,~~ |\nu_3\rangle\,,\qquad\qquad\qquad &\omega_3^{_{(+)}}&\simeq~\frac{\,\sin^2\!\theta_{_{\!S_3}}\,\delta M_3^2\,}{2E_\nu}\!\left(1+\frac{E^\text{res}_{\nu_3}}{E_\nu}\right)\label{effectiveMassHighE}
\end{alignat}
\end{subequations}

\begin{figure}
\centering
\makebox[\textwidth][c]{\includegraphics[width=1.0\textwidth]{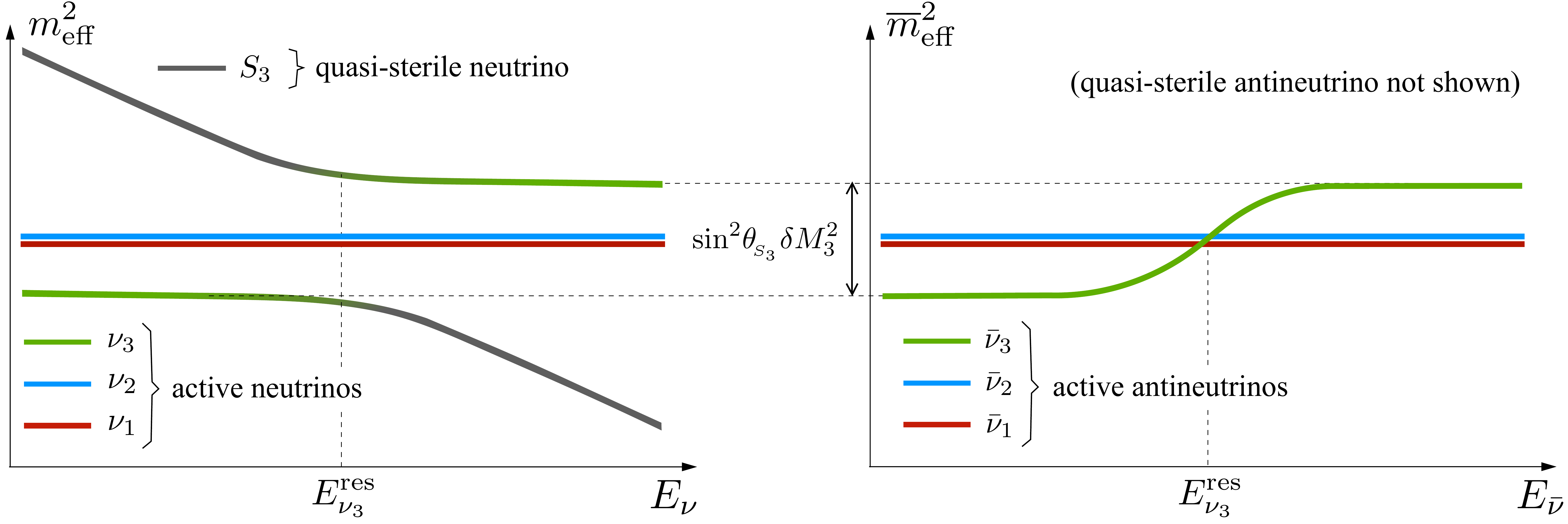}}
\caption{[\emph{For illustration purposes only --- not to scale\,}] Effective mass-squared eigenvalues of neutrino propagation eigenstates (left panel) and antineutrino propagation eigenstates (right panel) as a function of (anti-)neutrino energy in the quasi-sterile neutrino toy model. The eigenstates' flavor compositions are denoted by their color~shading.\label{CartoonToyModel}}
\end{figure}

Importantly, at high very neutrino energies $E_\nu\to \infty$, the active neutrino $\nu_3$ receives an effective contribution to its mass of $\delta m_3^2=\sin^2\!\theta_{_{\!S_3}}\,\delta M_3^2$ (note~(\ref{effectiveMassHighE})), an effect that had been previously noted in~\cite{Kopp-ml-2014fha,Denton-ml-2018dqq,Barenboim-ml-2019hso,Smirnov-ml-2021zgn}. Compatibility with observations from high energy atmospheric and accelerator neutrinos then enforces only two possibilities:
\begin{enumerate}[label=\alph*)]
\item \label{optionA}
The effective mass contribution to $\nu_3$ should be extremely suppressed so as to not make any difference, i.e., it should be of order of the current experimental uncertainty on the atmospheric mass-squared splitting. Quantitatively, this would imply
\begin{equation}
\sin^2\!\theta_{_{\!S_3}\,} \delta M_3^2~  \lesssim~  \mathcal{O}(10^{-4})~\text{eV}^2\,.
\end{equation}
\item \label{optionB}
In vacuum, the active neutrino mass hierarchy is inverted, and at high neutrino energies $E_\nu\gg E^\text{res}_{\nu_3}$ the effective mass contribution to $\nu_3$ due to matter effects flips the neutrino mass ordering to a normal hierarchy. This would require
\begin{align}\label{FlipRelation}
\sin^2\!\theta_{_{\!S_3}\,} \delta M_3^2 ~&\simeq~ 2 \, \left|\delta m^2_{32}\right| \\ &\sim~  (5\;\pm\;0.5)\times 10^{-3}~\text{eV}^2\,. \nonumber
\end{align}
\end{enumerate}

Option~(\ref{optionA} above would preclude any detectable effect at MiniBooNE, and therefore, while perfectly compatible with experimental constraints, it is not relevant for this minimal model case study. Which leaves us with~(\ref{optionB} as the only option of interest.

\end{itemize}

Figures~\ref{Theta3m} and~\ref{CartoonToyModel} provide a clear illustration of the three energy regimes discussed, as well as the transitions between them. In particular, figure~\ref{CartoonToyModel} (left panel) depicts how the composition and eigenvalues of the propagation eigenstates change as the neutrino energy crosses the resonant region set by $E^\text{res}_{\nu_3}$.

\subsection{Neutrino oscillation probabilities}
\label{OscProbSec}

We can now proceed to derive various appearance and disappearance oscillation probabilities. The first step is to obtain the flavor eigenstates as a function of the propagation eigenstates~in~(\ref{matterEigen}).

Inverting the relations in~(\ref{matterEigen3}) and~(\ref{matterEigen4}), we have:
\begin{subequations}
\begin{alignat}{3}
&|\nu_3\rangle~&&=\,~~\cos\theta_{_{\!S_3}}^{_{(m)}}\,|N_{3,\, m}^{_{(-)}}\rangle ~+~ \sin\theta_{_{\!S_3}}^{_{(m)}}\,|N_{3,\, m}^{_{(+)}}\rangle \label{nu3Expansion} \\[7pt]
&|S_3\rangle~&&=\,-\sin\theta_{_{\!S_3}}^{_{(m)}}\,|N_{3,\, m}^{_{(-)}}\rangle ~+~\cos\theta_{_{\!S_3}}^{_{(m)}}\,|N_{3,\, m}^{_{(+)}}\rangle \label{S3Expansion}
\end{alignat}
\end{subequations}

Using~(\ref{vacuumEigenstates}) and~(\ref{nu3Expansion}), the muon-neutrino flavor is then given by:
 \begin{align}
|\nu_\mu\rangle &= U^*_{\mu 1}\,|\nu_1\rangle  +  U^*_{\mu 2}\,|\nu_2\rangle  +  U^*_{\mu 3}\,|\nu_3\rangle  \\[7pt]
 &= U^*_{\mu 1}\,|\nu_1\rangle  +  U^*_{\mu 2}\,|\nu_2\rangle  +  U^*_{\mu 3\,}\cos\theta_{_{\!S_3}}^{_{(m)}}\,|N_{3,\, m}^{_{(-)}}\rangle  +  U^*_{\mu 3\,}\sin\theta_{_{\!S_3}}^{_{(m)}}\,|N_{3,\, m}^{_{(+)}}\rangle\,, \qquad \nonumber
\end{align}
which, after propagating through a distance $L$, is time-evolved to:
 \begin{align}\label{NuMUevolvedToy}
|\nu_\mu(L)\rangle &= U^*_{\mu 1}\,e^{-i\,\omega_1 L}\;|\nu_1\rangle  +  U^*_{\mu 2}\,e^{-i\,\omega_2 L}\;|\nu_2\rangle  \\[4pt]
&\quad +  U^*_{\mu 3\,}\cos\theta_{_{\!S_3}}^{_{(m)}}\,e^{-i\,\omega_3^{_{(-)}} L}\;|N_{3,\, m}^{_{(-)}}\rangle  +  U^*_{\mu 3\,}\sin\theta_{_{\!S_3}}^{_{(m)}}\,e^{-i\,\omega_3^{_{(+)}} L}\;|N_{3,\, m}^{_{(+)}}\rangle\,. \qquad \nonumber
\end{align}

Likewise, the electron-neutrino flavor is given by:
 \begin{align}
|\nu_e\rangle &= U^*_{e 1}\,|\nu_1\rangle  +  U^*_{e 2}\,|\nu_2\rangle  +  U^*_{e 3}\,|\nu_3\rangle  \\[7pt]
&= U^*_{e 1}\,|\nu_1\rangle  +  U^*_{e 2}\,|\nu_2\rangle  +  U^*_{e 3\,}\cos\theta_{_{\!S_3}}^{_{(m)}}\,|N_{3,\, m}^{_{(-)}}\rangle  +  U^*_{e 3\,}\sin\theta_{_{\!S_3}}^{_{(m)}}\,|N_{3,\, m}^{_{(+)}}\rangle\,, \qquad \nonumber
\end{align}
and through propagation evolves to:
 \begin{align}\label{NuEevolvedToy}
|\nu_e(L)\rangle &= U^*_{e 1}\,e^{-i\,\omega_1 L}\;|\nu_1\rangle  +  U^*_{e 2}\,e^{-i\,\omega_2 L}\;|\nu_2\rangle  \\[4pt]
&\quad +  U^*_{e 3\,}\cos\theta_{_{\!S_3}}^{_{(m)}}\,e^{-i\,\omega_3^{_{(-)}} L}\;|N_{3,\, m}^{_{(-)}}\rangle  +  U^*_{e 3\,}\sin\theta_{_{\!S_3}}^{_{(m)}}\,e^{-i\,\omega_3^{_{(+)}} L}\;|N_{3,\, m}^{_{(+)}}\rangle\,. \qquad \nonumber
\end{align}

We note that the expressions~(\ref{NuMUevolvedToy}) and~(\ref{NuEevolvedToy}) above are valid for neutrino propagation through a medium with constant density and composition. In cases where the medium density/composition changes as the neutrino propagates, so does the Hamiltonian change. In such cases,~(\ref{NuMUevolvedToy}) and~(\ref{NuEevolvedToy}) have to replaced with the appropriate states propagated under time-ordered Hamiltonian evolution.

The $\nu_\mu\to S_3$ oscillation probability is then given by:
\begin{align}
\label{muDis}
\text{P}(\nu_\mu\rightarrow S_3) &\equiv |\langle S_3|\nu_\mu(L)\rangle|^2\nonumber\\ &= |U_{\mu 3}|^2~\sin^22\theta_{_{\!S_3}}^{_{(m)}}~\sin^2\!\left(\frac{C_3\, \delta M_3^2}{4E_\nu}L\right)\,,
\end{align}
and the $\nu_e\to S_3$ oscillation probability by:
\begin{align}
\label{eDis}
\text{P}(\nu_e\rightarrow S_3) &\equiv |\langle S_3|\nu_e(L)\rangle|^2\nonumber\\ &= |U_{e 3}|^2~\sin^22\theta_{_{\!S_3}}^{_{(m)}}~\sin^2\!\left(\frac{C_3\,\delta M_3^2}{4E_\nu}L\right).
\end{align}

Finally, the $\nu_\mu\rightarrow\nu_e$ appearance probability is given by:
\begin{align}
\label{exactApp}
\text{P}(\nu_\mu\rightarrow\nu_e) &= |\langle\nu_e|\nu_\mu(L)\rangle|^2 \\[4pt] &= \Big|\,U_{e1}\,U^*_{\mu1}\,e^{-i\,\omega_1 L} + U_{e2}\,U^*_{\mu2}\,e^{-i\,\omega_2 L} \nonumber\\
&\qquad + U_{e3}\,U^*_{\mu3}\,\cos^2\!\theta_{_{\!S_3}}^{_{(m)}}\,e^{-i\,\omega_3^{_{(-)}}L} + U_{e3}\,U^*_{\mu3}\,\sin^2\!\theta_{_{\!S_3}}^{_{(m)}}\,e^{-i\,\omega_3^{_{(+)}}L}\,\Big|^2\,. \nonumber
\end{align}

In MiniBooNE, the baseline-to-energy ratio is in the range of $L/E_\nu \sim \mathcal{O}(1-10)\text{ eV}^{-2}$. Therefore, when computing the $\nu_e$ appearance probabilities at MiniBooNE, we can ignore the difference in oscillation phases between the two `solar' neutrino states $\nu_1$ and $\nu_2$, since $(\omega_2L-\omega_1L)\,|_\text{\tiny MB} \sim \mathcal{O}(10^{-5}-10^{-4})$. In this case we have
\begin{equation}
\label{MBApp}
\text{P}(\nu_\mu\rightarrow\nu_e)\Big|_\text{MB} \simeq |U_{e3}|^2\,|U_{\mu3}|^2\,\left| - e^{-i\,\omega_2\,L}+\cos^2\!\theta_{_{\!S_3}}^{_{(m)}}\,e^{-i\,\omega_3^{_{(-)}}L}+\sin^2\!\theta_{_{\!S_3}}^{_{(m)}}\,e^{-i\,\omega_3^{_{(+)}}L}\,\right|^2.
\end{equation}

\vspace{10pt}
We can look at three energy regimes of MiniBooNE's  appearance probability in~(\ref{MBApp}):

\vspace{10pt}
\noindent{\bf For low energy neutrinos}, {$ E_\nu\ll E^\text{res}_{\nu_3}$,} or alternatively, at negligible medium densities, $V_m\to 0$, the appearance probability~(\ref{MBApp}) reduces to the familiar form:
\begin{equation}
\text{P}(\nu_\mu\rightarrow\nu_e)\,\Big|_{E_\nu \to  0} \approx 4\;|U_{e3}|^2\,|U_{\mu3}|^2\,\sin^4\!\theta_{_{\!S_3}}~\sin^2\!\left(\frac{\delta M_3^2}{4E_\nu}L\right).
\end{equation}

\vspace{5pt}
\noindent{\bf At the resonance}, $E_\nu = E^\text{res}_{\nu_3}$,
\begin{eqnarray}\label{appProbRes}
\text{P}(\nu_\mu\rightarrow\nu_e)\,\Big|_{E^\text{res}_{\nu_3}} \approx 4\,|U_{e3}|^2|U_{\mu3}|^2\left[ \sin^2\!\left(\frac{\,2\,\delta m_{32}^2/\!\sin\theta_{_{\!S_3}}\,}{4\,E^\text{res}_{\nu_3}}L\right) -\frac{1}{4}\sin^2\!\left(\frac{\,4\,\delta m_{32}^2/\!\sin\theta_{_{\!S_3}}\,}{4\,E^\text{res}_{\nu_3}}L\right) \right],\nonumber\\
 \end{eqnarray}
where we have used~(\ref{FlipRelation}). Note that the $\nu_e$ appearance probability is maximized at the resonance, and, in the case of~(\ref{appProbRes}), it is upper-bounded by $4\,|U_{e3}|^2|U_{\mu3}|^2\sim(0.04-0.06)$. It peaks at:
\begin{equation}\label{appProbResMax}
\frac{\delta m_{32}^2}{\,\sin\theta_{_{\!S_3}}\,}\frac{L}{\,E^\text{res}_{\nu_3}\,} = (2n+1)\,\pi,\qquad n\in\mathbb{Z}\,.
\end{equation}
If we set $L$ in~(\ref{appProbResMax}) to MiniBooNE's baseline\footnote{The center of the MiniBooNE detector is located 541\,m from the production target; however, the baseline's first $ \sim $50\,m constitute the decay region~\cite{MiniBooNE-ml-2008paa}. Hence, the beam neutrinos propagate in matter through a distance of $ \sim $490\,m. In this specific case, the evolution of the initial $\nu_\mu$ state through the decay region can be neglected, and the MiniBooNE baseline can be set to $L|_\text{\tiny MB}=490$\;m.} of $L|_\text{\tiny MB}=490$\;m, and also set $n=0$ or $n=-1$, we can obtain the maximum value of $\big|{\sin\theta_{_{\!S_3}}\!}\big|$ that is able to saturate the appearance probability at the resonance. Choosing $E^\text{res}_{\nu_3}\big|_\text{\tiny MB}=300$\;MeV for concreteness, we have
\begin{equation}\label{Theta3max}
\big|{\sin\theta_{_{\!S_3}}^{\text{(max)}}}\big| = \frac{\,|\delta m_{32}^2|\,}{\pi}~\frac{L}{\,E^\text{res}_{\nu_3}\,}\Bigg|_\text{MB} \qquad \Rightarrow\qquad  \big|{\theta_{_{\!S_3}}^{\text{(max)}}}\big| \simeq 6.6\times 10^{-3}\,.
\end{equation}

\begin{figure}
\centering
\includegraphics[width=1\textwidth,trim= 0 0 40 0,clip]{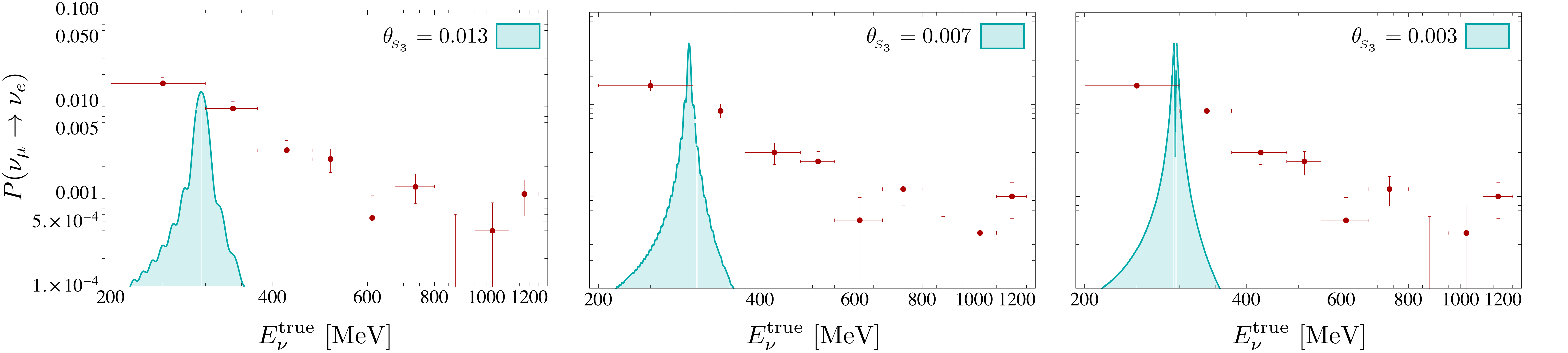}
\caption{{The $\nu_\mu\to\nu_e$ appearance probability at MiniBooNE, as a function of true neutrino energy, predicted by the toy model~(\ref{MBApp}) assuming $E^\text{res}_{\nu_3}\big|_\text{\tiny MB}=295$\;MeV}. Three different choices for the active-sterile vacuum mixing angle $\theta_{_{\!S_3}}$ are shown: below (left panel), at (middle panel), and above (right panel) the appearance probability saturation condition; see~(\ref{appProbResMax}) and~(\ref{Theta3max}). The red data points with error bars were taken from figure\,21 of~\cite{MiniBooNE-ml-2020pnu}; they assume $E_\nu^\text{true}\sim E_\nu^\text{\tiny QE}$, which is a very coarse approximation. As such, they should be interpreted as a \emph{qualitative} measure of MiniBooNE's anomalous appearance~probabilities.\label{ToyProbApp}}
\end{figure}

Figure~\ref{ToyProbApp} shows the toy model's $\nu_e$ appearance probability at MiniBooNE for $E^\text{res}_{\nu_3}\big|_\text{\tiny MB}=295$\;MeV and three different choices of $\big|\theta_{_{\!S_3}\!}\big|$. In the left panel $\big|\theta_{_{\!S_3}\!}\big| > \big|{\theta_{_{\!S_3}}^{\text{(max)}}}\big|$, and therefore the $\nu_e$ appearance probability is below the saturation condition in~(\ref{appProbResMax}). In the middle panel $\big|\theta_{_{\!S_3}\!}\big| \approx \big|{\theta_{_{\!S_3}}^{\text{(max)}}}\big|$; the appearance probability is saturated and displays a single sharp peak at the resonant energy. Finally, in the right panel $\big|\theta_{_{\!S_3}\!}\big| < \big|{\theta_{_{\!S_3}}^{\text{(max)}}}\big|$; the appearance probability is saturated but oscillates rapidly within the resonant region.

The span of the resonant region is nearly-independent of $\theta_{_{\!S_3}}$ and $E^\text{res}_{\nu_3}\big|_\text{\tiny MB}$.
As a rough approximation, we can consider the width of the resonant oscillations' envelope, $\Delta E^\text{res}$, whose magnitude is given by:
\begin{align}
\Delta E^\text{res}\big|_{_\text{\tiny MB}} &\sim 2~\big|{\tan 2\theta_{_{\!S_3}}^{\text{(max)}}}\big|~E^\text{res}_{\nu_3}\big|_{_\text{\tiny MB}}\nonumber\\
&\sim\frac{4}{\pi}~|\delta m_{32}^2|~L\big|_{_\text{\tiny MB}}\nonumber\\[3pt]
&\sim8\;\text{MeV}\,,
\end{align}
which is narrow  for MiniBooNE's detector resolution. Averaging the oscillation probability within the resonant range $\Delta E^\text{res}$ yields about $(40-45)\%$ of the peak value of $4\,|U_{e3}|^2|U_{\mu3}|^2$; i.e., P$(\nu_\mu\rightarrow\nu_e)\sim(1.6-2.7)\%$ in the resonant region. Nuclear effects and energy resolution preclude the MiniBooNE detector from resolving such a narrow peak in the energy distribution of $\nu_e$ events. Smearing this rate over the average bin size of 100\,MeV, we obtain an average appearance probability of roughly $~(0.3-0.5)\%$ within the $200-300$\;MeV range, which is still too small to account for the reported excess in appearance probability of $1.4-1.8$\% in this energy bin~\cite{MiniBooNE-ml-2020pnu}. This is compounded by the fact that this model predicts a negligible $\nu_e$ appearance probability in the regions outside the resonant region, including MiniBooNE's higher energy bins with reported excesses (see figure~\ref{ToyProbApp}). This constitutes one of three important reasons why this minimal model falls short of explaining the MiniBooNE excess. We will come back to this issue shortly.

\vspace{5pt}
\noindent{\bf For very high energy neutrinos}, $E_\nu \gg  E^\text{res}_{\nu_3}$, or alternatively, in very dense matter, $|\Delta V|\gg \delta M_3^2/E_\nu$,~(\ref{MBApp}) reduces to:
\begin{equation}\label{appProbHighE}
\text{P}(\nu_\mu\rightarrow\nu_e)\,\Big|_{E_\nu \gg  E^\text{res}_{\nu_3}} ~\approx~ 4\,|U_{e3}|^2|U_{\mu3}|^2\,\sin^2\!\left[\,\frac{\,\delta m_{32}^2\,(1+2\,E^\text{res}_{\nu_3}/E_\nu)\,}{4E_{\nu}}\,L\,\right].
\end{equation}
Note that~(\ref{appProbHighE}) recovers the SM limit of $\nu_\mu\to \nu_e$ oscillation probabilities at very high energies, since the effective mass-squared splitting controlling these oscillations, namely, $\delta m_{32}^2\,(1+2\,E^\text{res}_{\nu_3}/E_\nu)$, approaches the SM value of~$\delta m_{32}^2$~as~$E^\text{res}_{\nu_3}/E_\nu \to 0$. (As discussed in option~(\ref{optionB} of subsection~\ref{regimes}, this is because the condition in~(\ref{FlipRelation}) effectively flips the active neutrino in-medium mass hierarchy at high energies.)

Unfortunately, however, the active neutrino mass ordering does not relax to the SM expectation quickly enough at high energies, leading to strong tensions with measurements from long-baseline neutrino experiments. Consider, for instance, T2K's measurement of $\delta m_{32}^2$ in neutrino mode. The average matter density along the Tokai-to-Kamioka baseline is $\rho|_\text{\tiny T2K}=2.6$\;g/cm$^3$~\cite{Hagiwara-ml-2011kw}; therefore, the resonant energy at T2K is related to the resonant energy at MiniBooNE through:
\begin{align}\label{EresT2K}
E^\text{res}_{\nu_3}\big|_{_\text{\tiny T2K}}&\approx E^\text{res}_{\nu_3}\big|_{_\text{\tiny MB}}~~\frac{\;\rho|_{_\text{\tiny MB}}}{\;\rho|_{_\text{\tiny T2K}}}\nonumber\\ &\sim E^\text{res}_{\nu_3}\big|_{_\text{\tiny MB}}~~\frac{[1.6-2.2] \text{ g/cm}^{\! 3}}{\;2.6  \text{ g/cm}^{\! 3}\;}\,.
\end{align}
Above, the estimated range of soil densities along MiniBooNE's baseline, $\rho|_{_\text{\tiny MB}}\approx[1.6-2.2] \text{ g/cm}^{\! 3}$, is based on the assumption that the soil's composition is dominated by silty clay~\cite{MBsoil}. Then, for the benchmark choice of $E^\text{res}_{\nu_3}\big|_\text{\tiny MB}=300$\;MeV,~(\ref{EresT2K}) gives $E^\text{res}_{\nu_3}\big|_\text{\tiny T2K}\sim [185-254]$\;MeV\@. For the remainder of this paper, we will benchmark the soil density along MiniBooNE's baseline to $\rho|_{_\text{\tiny MB}}\approx 1.6 \text{ g/cm}^{\! 3}$, and therefore we will use $E^\text{res}_{\nu_3}\big|_\text{\tiny T2K}\sim 185$\;MeV\@.  With this information, we can consider T2K's determination of $\delta m_{32}^2$, which is dominated by its sensitivity to the dip in the $\nu_\mu$ survival probability at around $E_{\nu_\mu}^{\text{\,dip}}\big|_\text{\tiny T2K}\sim 600$\;MeV\@. Using~(\ref{appProbHighE}), the effective $\delta m_{32 \text{,\,eff}}^2$ predicted at T2K in this energy range would be:
\begin{align}\label{EffM32sqT2K}
\delta m_{32 \text{,\,eff\,}}^2\big|_{_\text{T2K}} &\sim |\delta m_{32}^2|\big|_{_\text{SM}}\times \left(\!1+2\,\frac{\, E^\text{res}_{\nu_3} \,}{\;E_{\nu_\mu}^{\text{\,dip}}\;} \right)\!\Bigg|_{_\text{T2K}}\nonumber\\ &\sim  1.62\times |\delta m_{32}^2|\big|_{_\text{SM}}\,,
\end{align}
which is in significant tension with neutrino data from T2K~\cite{bibT2K-ml-2021xwb}, whose error on its most recent determination of $\delta m_{32}^2$ is only about 4\%. This constitutes the second of three main reasons why this minimal model is phenomenologically unviable.

\begin{figure}
\centering
\includegraphics[width=0.7\textwidth,trim= 0 0 50 0,clip]{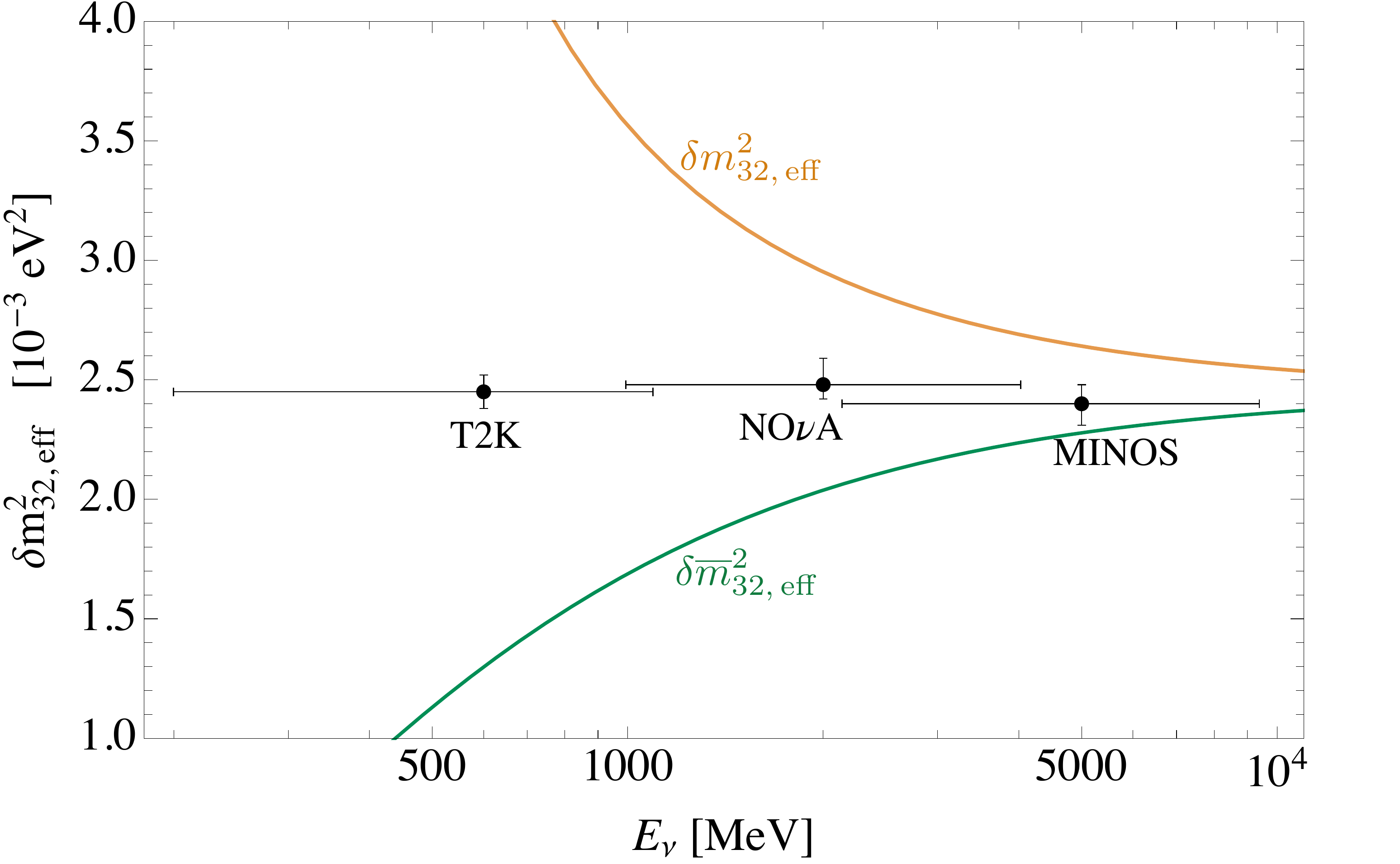}
\caption{The effective `atmospheric' mass-squared splitting predicted by the toy model for both neutrinos ($\delta m_{32 \text{,\,eff}}^2$) and antineutrinos ($\delta \overline{m}_{32 \text{,\,eff}}^2$) as a function of (anti-)neutrino energy, assuming $\theta_{_{\!S_3}}=0.003$, $E^\text{res}_{\nu_3}\big|_\text{\tiny MB}\sim 300$\;MeV, and $E^\text{res}_{\nu_3}\big|_\text{\tiny T2K}\simeq E^\text{res}_{\nu_3}\big|_\text{\tiny NO$\nu$A}\simeq E^\text{res}_{\nu_3}\big|_\text{\tiny MINOS}\sim 185$\;MeV\@. The data points with error bars denote recent combined measurements of $\delta m_{32}^2$ by long-baseline accelerator experiments~\cite{bibT2K-ml-2018rhz,MINOS-ml-2020llm,NOvA-ml-2021nfi}.\label{ToyMeff}}
\end{figure}
Figure~\ref{ToyMeff} shows this toy model's prediction for the effective $\delta m_{32 \text{,\,eff}}^2$ as a function of neutrino energy in long-baseline neutrino experiments such as T2K~\cite{bibT2K-ml-2018rhz}, MINOS~\cite{MINOS-ml-2020llm}, and NO$\nu$A~\cite{NOvA-ml-2021nfi}. Here, we take $\theta_{_{\!S_3}}=0.003$ and make the same simplified assumption that the average matter density through which neutrinos propagate in these long-baseline experiments results in a long-baseline resonant energy of $E^\text{res}_{\nu_3}\big|_\text{\tiny T2K}\sim 185$\;MeV (see discussion around~(\ref{EresT2K})). It is clear from figure~\ref{ToyMeff} that severe tensions between data and the toy model prediction for $\delta m_{32 \text{,\,eff}}^2$  persist even at much higher neutrino energies than the range probed by T2K.

\subsection{Matter effects on antineutrinos}
\label{ToyAntiNU}
We now briefly discuss the consequences of the minimal model for antineutrinos. The sign of the matter potential, $\Delta V$, was chosen such that antineutrinos do not experience resonant oscillations. Furthermore, the maximum allowed strength of the $\bar\nu_3-\bar{S}_3$ mixing angle, given in~(\ref{Theta3max}), is too small to yield a significant amplitude for non-resonant $\bar\nu_3-\bar{S}_3$ oscillations (see also figure~\ref{Theta3m}). Consequently, the $\bar\nu_\mu\to\bar\nu_e$ appearance probability predicted by this minimal model is negligible, and therefore cannot account for the excesses of $\bar\nu_e$-like events reported by LSND and MiniBooNE\@. This is the third main reason why this minimal model is phenomenologically unviable.

Additionally, there is another shortcoming of the minimal model which is related to the second reason discussed at the end of subsection~\ref{OscProbSec}. Before we state it, we need to highlight how these BSM matter effects alter the in-medium oscillations of antineutrinos. Their propagation eigenstates are given by analogous expressions to~(\ref{matterEigen}) and~(\ref{effMixAngle}) with~$|\Delta V|\to-|\Delta V|$:
\begin{subequations}\label{AntiNuMatterEigen}
\begin{alignat}{3}
&\!\!\!\!\!\!\!\! |\bar{N}_{1,\,m}\rangle~=~|\bar{\nu}_1\rangle\,, &\bar{\omega}_1~~&=~\frac{\delta m^2_{13}}{2 E_{\bar\nu}} \label{AntiNuEigen1}\\
&\!\!\!\!\!\!\!\! |\bar{N}_{2,\,m}\rangle~=~|\bar{\nu}_2\rangle\,, &\bar{\omega}_2~~&=~\frac{\delta m^2_{23}}{2 E_{\bar\nu}} \label{AntiNuEigen2}\\
&\!\!\!\!\!\!\!\! |\bar{N}_{3,\, m}^{_{(-)}}\rangle~=-\sin\bar\theta_{_{\!S_3}}^{_{\,(m)}}|\bar{S}_3\rangle \,+\, \cos\bar\theta_{_{\!S_3}}^{_{\,(m)}}|\bar{\nu}_3\rangle\,,~~~~ &\bar{\omega}_3^{_{(-)}} &=~\frac{\delta M_3^2}{4E_{\bar\nu}}\left(\!1-\overline{C}_3+\frac{E_{\bar\nu}}{E^\text{res}_{\nu_3}} \cos 2\theta_{_{\!S_3}}\!\right)\!\!\! \label{AntiNuEigen3}\\
&\!\!\!\!\!\!\!\!  |\bar{N}_{3,\, m}^{_{(+)}}\rangle~=~~\cos\bar\theta_{_{\!S_3}}^{_{\,(m)}}|\bar{S}_3\rangle \,+\, \sin\bar\theta_{_{\!S_3}}^{_{\,(m)}}|\bar{\nu}_3\rangle\,,~~~~ &\bar{\omega}_3^{_{(+)}} &=~\frac{\delta M_3^2}{4E_{\bar\nu}}\left(\!1+\overline{C}_3+\frac{E_{\bar\nu}}{E^\text{res}_{\nu_3}} \cos 2\theta_{_{\!S_3}}\!\right)\!\!\! \label{AntiNuEigen4}
\end{alignat}
\end{subequations}
where we have recast $|\Delta V|$ in terms of $E^\text{res}_{\nu_3}$ using~(\ref{Eres3}). Furthermore, $\overline{C}_3$ and the effective $\bar{\nu}_3-\bar{S}_3$ mixing angle in matter, $\bar\theta_{_{\!S_3}}^{_{\,(m)}}$, are defined through:
\begin{equation}\label{AntiNuEffMixAngle}
\sin2\bar\theta_{_{\!S_3}}^{_{\,(m)}} = \frac{\sin2\theta_{_{\!S_3}}}{\overline{C}_3^{}} \equiv \frac{\sin2\theta_{_{\!S_3}}}{\;\sqrt{~\big(\!\sin2\theta_{_{\!S_{3\!}}}\big)^2 + \big(\!\cos2\theta_{_{\!S_{3\!}}}\big)^2\left(1+\frac{E_{\bar\nu}}{\,E^\text{res}_{\nu_3}\,}\right)^2\;}}\,.
\end{equation}

Figures~\ref{Theta3m} and~\ref{CartoonToyModel} (right panel) illustrate, respectively, the in-medium $\bar{\nu}_3-\bar{S}_3$ mixing angle as a function of antineutrino energy, and the flavor composition and eigenvalues of the in-medium propagation eigenstates as a function of antineutrino energy.

We can proceed to examine two energy regimes of experimental interest: (i) the low energy region $E_{\bar\nu}\sim\mathcal{O}(1-10)\,$MeV, relevant for reactor experiments; and (ii) the high energy region $E_{\bar\nu}\;\gsim\; (0.2-10)\,$GeV, relevant for long-baseline accelerator experiments. Firstly, we note that in this minimal model $\bar{\nu}_3-\bar{S}_3$ mixing is always suppressed, and the active neutrino state $\bar{\nu}_3$ is mostly composed of $\bar{N}_{3}^{_{(-)}}$. Therefore, the effective antineutrino mass splitting $\delta \overline{m}_{3i \text{,\,eff\,}}^2$ ($i=1,2$) can be approximated by:
\begin{eqnarray}\label{EffM31sqLowE}
\delta \overline{m}_{3i \text{,\,eff\,}}^2\Big|_{\bar\nu}~&\simeq&~~~2\,E_{\bar\nu}\,\left(\bar{\omega}_3^{_{(-)}}-\bar{\omega}_i \right)\nonumber\\
&~\rightarrow&~\begin{dcases}
  	    -\,|\delta m_{3i}^2|\,\left(1-2\cos 2\theta_{_{\!S_3}} \frac{E_{\bar\nu}}{E^\text{res}_{\nu_3}} \right)&\text{for }E_{\bar\nu}\,\ll\,E^\text{res}_{\nu_3}\,,\\
  	    ~~~|\delta m_{3i}^2|\,\left(1-2\,\frac{E^\text{res}_{\nu_3}}{E_{\bar\nu}}\right) &\text{for }E_{\bar\nu}\,\gg\,E^\text{res}_{\nu_3}\,.
   	 \end{dcases}
\end{eqnarray}
Interestingly, as it is clear from~(\ref{EffM31sqLowE}), the antineutrino mass ordering in this minimal model displays the same `hierarchy-flipping' pattern of the active neutrinos, with an inverted mass hierarchy in the asymptotically low energy limit and a normal in-medium mass hierarchy in the asymptotically high energy limit. This effect is illustrated in figure~\ref{CartoonToyModel} (right panel).

Reactor-based antineutrino experiments such as Daya Bay~\cite{DayaBay-ml-2018yms} and RENO~\cite{RENO-ml-2018dro} have measured $|\delta m^2_{31}|$ with $\sim\!5\%$ precision at $2\sigma$\;C.L. within the energy range $E_{\bar\nu}\sim\mathcal{O}(1-10)\,$MeV\@. From~(\ref{EffM31sqLowE}), we can check that the minimal model is within $2\sigma$-compatibility with these measurements, predicting
\begin{equation}\label{EffM31sqReactor}
\delta \overline{m}_{31 \text{,\,eff\,}}^2\big|_{\text{reactor }\bar\nu} ~\simeq~  - |\delta m_{31}^2|\big|_{_\text{SM}}\,\left[1 - 0.02\times \!\bigg(\!\frac{E_{\bar\nu}}{\,3\,\text{MeV}\,}\!\bigg) \right].
\end{equation}

The minimal model's compatibility with $\bar\nu$-data is not as good in the high energy limit, however. Several long-baseline accelerator experiments, such as K2K, T2K, MINOS, and NO$\nu$A have also measured the active mass-squared splittings and mixing angles in antineutrino mode, albeit with less precision that in neutrino mode. Taking T2K again as an example, we can check the minimal model's prediction for the effective $\delta \overline{m}_{32 \text{,\,eff}}^2$ that would be observed at T2K in antineutrino mode:
\begin{align}\label{EffM32sqT2KantiNU}
\delta \overline{m}_{32 \text{,\,eff\,}}^2\big|_{_{\text{T2K,}\,\bar\nu}} &~\sim~ |\delta m_{32}^2|\big|_{_\text{SM}}\times \left(\!1-2\,\frac{\, E^\text{res}_{\nu_3} \,}{\;E_{\bar\nu_\mu}^{\text{\,dip}}\;} \right)\!\Bigg|_{_\text{T2K}}\nonumber\\ &~\sim~  0.38\times |\delta m_{32}^2|\big|_{_\text{SM}}\,,
\end{align}
which is in significant tension with T2K's most recent determination of $\delta m_{32}^2$ in antineutrino mode~\cite{Berns-ml-2021iss}, which had an error of $\sim$8\%. This tension is analogous to the one~(\ref{EffM32sqT2K}) for neutrino mode, with the main difference being that the latter overpredicts $|\delta m_{32}^2|$, whereas the former underpredicts it. This is clearly illustrated in figure~\ref{ToyMeff}.

\section{Beyond the minimal model: a viable explanation for the MiniBooNE excess}
\label{FullModel}

We begin this section by recapping the phenomenological shortcomings of the minimal model just discussed:
\begin{enumerate}

\item \label{Problem1} \emph{The strength of the resonant $\nu_\mu\to\nu_e$ oscillations in the minimal model is insufficient to account for the full low energy $\nu_e$-like excess at MiniBooNE.}

There are two reasons for this. First, the resonant energy window is very narrow. Second, the active neutrino experiencing resonant oscillations, $\nu_3$, has a suppressed $\nu_e$ component, thereby suppressing the resonant amplitude by a factor of $|U_{e3}|^2|U_{\mu3}|^2\sim(0.01-0.015)$, as shown in~(\ref{MBApp}) and figure~\ref{ToyProbApp}.

\item \label{Problem2} \emph{The effective `atmospheric' splitting $\delta m_{32}^2$ predicted at long-baseline experiments is in strong tension with data.}

This is because, in the high energy regime $E_\nu\gg E^\text{res}_{\nu_3}$, the active state that mixes with the quasi-sterile state (which is $\nunubar_{\!3}$ in the minimal model) receives a residual contribution to its in-medium effective mass given by~(\ref{effectiveMassHighE}) for $\nu_3$ and~(\ref{AntiNuEigen3}) for $\bar\nu_3$. Hence, in order to preserve the expected active neutrino mass-squared splittings at high energies, the parameters of the minimal model have to be adjusted to effectively `flip' the mass hierarchy at $E_\nu\gg E^\text{res}_{\nu_3}$ according to~(\ref{FlipRelation}). However, this hierarchy-flip does not happen `instantaneously' as $E_\nu$ crosses the resonant region --- the effective mass-squared splitting $\delta m_{32 \text{,\,eff\,}}^2$ does not fully relax to the SM value until high enough energies, well above the energy range probed by, e.g., T2K, NO$\nu$A, and MINOS\@. As a result, the predicted $\delta m_{32 \text{,\,eff\,}}^2$ is in tension with accelerator-based measurements of $\delta m_{32}^2$, as shown in~(\ref{EffM32sqT2K}),~(\ref{EffM32sqT2KantiNU}), and figure~\ref{ToyMeff}.

\item \label{Problem3} \emph{The minimal model cannot address the $\bar\nu_e$-like excesses at LSND and MiniBooNE, nor the $\nu_e$-like excess at MiniBooNE above $E_\nu\gsim400$}\;MeV.

This issue is related to~\ref{Problem2} above. Since $\nu_3-S_3$ mixing has been constrained by the `hierarchy-flip' relation in~(\ref{FlipRelation}), this model predicts negligible appearance probabilities for $\bar\nu_\nu\to\bar\nu_e$ oscillations over all energy ranges, as well as for $\nu_\nu\to\nu_e$ oscillations outside the resonant region (see figure~\ref{ToyProbApp}).
\end{enumerate}

The reasons behind these shortcomings, as summarized above, offer clues as to how the minimal model could be augmented and modified to overcome phenomenological tensions.

In particular, shortcoming~\ref{Problem1} above could be addressed if $\nu_1$ and/or $\nu_2$ --- which have larger $\nu_e$ components --- were involved in the resonant oscillation mechanism. However, this needs to be done in such a way that the effective mass spectrum of active neutrinos is not significantly altered at high energies. In other words, the incorporation of resonance effects for $\nu_{1,2}$ in order to address shortcoming~\ref{Problem1} must simultaneously address shortcoming~\ref{Problem2}.

The most straightforward way to achieve this goal is to augment the minimal model by introducing two additional quasi-sterile species, $S_1$ and $S_2$, which will mix with $\nu_1$ and $\nu_2$, respectively. This will result in two new resonant energies, $E^\text{res}_{\nu_1}$ and $E^\text{res}_{\nu_2}$, which must fall within the low energy window of interest for MiniBooNE\@. These additional resonances will be weighted by factors of $|U_{e1}|^2|U_{\mu1}|^2$ and $|U_{e2}|^2|U_{\mu2}|^2$, respectively, and therefore will allow for much larger resonant appearance probabilities than the minimal model. Furthermore, the parameters of this augmented model must be arranged such that the effective shift in the active neutrino masses is the same for $\nu_1$, $\nu_2$, and $\nu_3$, so that the in-medium active mass hierarchy is preserved at high energies $E_\nu \gg E^\text{res}_{\nu_{1,2,3}}$. It turns out that this latter requirement will be sufficient to address shortcoming~\ref{Problem2} as well, as we will detail shortly.

Unfortunately, the model's augmentation by two new quasi-sterile states $S_1$ and $S_2$ is still insufficient to address shortcoming~\ref{Problem3}. As we shall see, this is because the three resonant regions will remain narrow, and $\nunubar_{\!i}-\SSbar_{\!i}$ oscillations will remain suppressed away from the resonant region. Therefore, a fully viable model would still need an additional source of non-resonant active-to-sterile oscillations. The simplest option is to introduce a new `vanilla' sterile neutrino that is neutral under the dark sector gauge interactions, and therefore does not experience a matter potential. This vanilla sterile state will mix with a linear combination of active neutrinos to generate an off-resonance appearance probability, and, importantly, it will not qualitatively interfere with the resonant mechanism. This ordinary `3+1'-like mechanism is well-known for providing an excellent fit to the LSND oscillation data, as well as to the MiniBooNE oscillation data for antineutrinos and mid-energy neutrinos. If one wishes, the vanilla `3+1' parameters could also be adjusted to fit $\nunubar_{\!e}$ disappearance data, such as the Gallium and/or reactor anomalies, in the spirit of global fits.

Our goal in this study, however, is not to perform global fits of this model, much less re-fit the vanilla `3+1' parameters to all existing neutrino data (something that has already been done \emph{ad nauseam} in the literature; see, e.g.,~\cite{Kopp-ml-2013vaa,Gariazzo-ml-2017fdh,Dentler-ml-2018sju,Diaz-ml-2019fwt,Moulai-ml-2019gpi,Berryman-ml-2021yan,Dasgupta-ml-2021ies,Gonzalez-Garcia-ml-2021dve}). Instead, we will benchmark the vanilla `3+1' parameters to yield a good fit to the LSND, MiniBooNE $\bar\nu$-mode, and Gallium anomalies, and focus our attention on deconstructing the parameter space of the resonant mechanism.

\subsection{Full model --- part I: the triple-resonance mechanism}

For pedagogical purposes, we begin by describing the resonant mechanism of the full model with the `3+1' non-resonant effects tuned off. In the next subsection, we will then introduce the `vanilla' sterile state to  complete the description of the full model.

As stated in the previous subsection, the full model will contain three quasi-sterile neutrinos, $S_1$, $S_2$, and $S_3$, all of which experience the same effective matter potential in~(\ref{Vm}). Each of the $S_i$ will mix exclusively with $\nu_i$ with a vacuum mixing angle $\theta_{_{\!S_i}}$, in a manner analogous to the $\nu_3-S_3$ mixing discussed in section~\ref{toymodel}.

The effective matter Hamiltonian in the basis $B=\{|\nu_1\rangle,|S_1\rangle,|\nu_2\rangle,|S_2\rangle,|\nu_3\rangle,|S_3\rangle\}$ can then be written in block-diagonal form as:
\begin{equation}\label{fullH}
\!\!\!\!\widehat{H}_{B,\text{ matter}} =
\begin{pmatrix}
\widehat{H}_1^{\,(2\times2)} &  &  \\
 & \widehat{H}_2^{\,(2\times2)} &  \\
 &  & \widehat{H}_3^{\,(2\times2)}
\end{pmatrix}
\end{equation}
where each block $\widehat{H}_i^{\,(2\times2)}$ acts on the subspace $\{|\nu_i\rangle,|S_i\rangle\}$ and is defined by:
\begin{align}\label{subH}
\widehat{H}_i = \widehat{H}\!\left[\nu_i,S_i\right] &\equiv
\begin{pmatrix}
 ~~\cos\theta_{_{\!S_i}} & \sin\theta_{_{\!S_i}} \\
-\sin\theta_{_{\!S_i}} & \cos\theta_{_{\!S_i}}
\end{pmatrix}\cdot
\begin{pmatrix}
\frac{m_i^2}{2E_\nu} & ~0 \\
0~ & \frac{M_i^2}{2E_\nu}
\end{pmatrix}\cdot
\begin{pmatrix}
\,\cos\theta_{_{\!S_i}} & -\sin\theta_{_{\!S_i}} \\
 \,\sin\theta_{_{\!S_i}} & ~~\cos\theta_{_{\!S_i}}
\end{pmatrix} -
\begin{pmatrix}
~0~ & 0 \\
~0~ & |\Delta V| \!
\end{pmatrix}\nonumber\\[8pt]
& = \frac{m_i^2}{2E_\nu}\begin{pmatrix}
\,1 ~&~0\,\, \\
\,0 ~& ~0\,\,
\end{pmatrix}
 \,+~ \frac{\delta M_i^2}{4E_\nu}\left(1-\frac{E_\nu}{\,E^\text{res}_{\nu_i}\,}\cos2\theta_{_{\!S_i}}\!\right) \begin{pmatrix}
\,1 ~&~0\,\, \\
\,0 ~& ~1\,\,
\end{pmatrix}
\\
&\quad + \frac{\delta M_i^2}{4E_\nu}
\begin{pmatrix}
-\cos2\theta_{_{\!S_i}}\!\Big(\!1-\frac{E_\nu}{\,E^\text{res}_{\nu_i}\,}\!\Big)~ &~ \sin2\theta_{_{\!S_i}} \\
~~\sin2\theta_{_{\!S_i}}~ &~ \cos2\theta_{_{\!S_i}}\!\Big(1-\frac{E_\nu}{\,E^\text{res}_{\nu_i}\,}\Big)\nonumber
\end{pmatrix}.
\end{align}
Above, $\delta M_i^2\equiv M_i^2-m_i^2$, and $M_i$ is the (vacuum) mass of the heavier eigenstate, which is composed predominantly of the sterile state $S_i$. Note that in the last equality of~(\ref{subH}) we have traded the matter potential $|\Delta V|$ for $E^\text{res}_{\nu_i}$ using:
\begin{equation}\label{tripleResDeltaV}
|\Delta V| ~=~ \frac{\delta M_1^2}{\,2\,E^\text{res}_{\nu_1}\,}\cos2\theta_{_{\!S_1}} = \frac{\delta M_2^2}{\,2\,E^\text{res}_{\nu_2}\,}\cos2\theta_{_{\!S_2}} = \frac{\delta M_3^2}{\,2\,E^\text{res}_{\nu_3}\,}\cos2\theta_{_{\!S_3}}\,.
\end{equation}

In analogy with~(\ref{matterEigen}), we can proceed to diagonalize~(\ref{subH}) to obtain the propagation eigenstates for $i=1,2,3$:
\begin{subequations}\label{MatterEigeni}
\begin{alignat}{3}
|{N}_{i,\, m}^{_{(-)}}\rangle &= -\sin\theta_{_{\!S_i}}^{_{(m)}}|{S}_i\rangle  +  \cos\theta_{_{\!S_i}}^{_{(m)}}|{\nu}_i\rangle\,,~~~~~~ &{\omega}_i^{_{(-)}} &= \frac{m_i^2}{2E_{\nu}} + \frac{\delta M_i^2}{4E_{\nu}}\left(\!1-{C}_i-\frac{E_{\nu}}{E^\text{res}_{\nu_i}} \cos 2\theta_{_{\!S_i}}\!\right)\!\!\! \label{NuEigenLi}\\[5pt]
 |{N}_{i,\, m}^{_{(+)}}\rangle &= ~~\cos\theta_{_{\!S_i}}^{_{(m)}}|{S}_i\rangle  +  \sin\theta_{_{\!S_i}}^{_{(m)}}|{\nu}_i\rangle\,,~~~~~~ &{\omega}_i^{_{(+)}} &= \frac{m_i^2}{2E_{\nu}} + \frac{\delta M_i^2}{4E_{\nu}}\left(\!1+{C}_i-\frac{E_{\nu}}{E^\text{res}_{\nu_i}} \cos 2\theta_{_{\!S_i}}\!\right)\!\!\! \label{NuEigenHi}
\end{alignat}
\end{subequations}
where, as usual, ${C}_i$ and the effective ${\nu}_i-{S}_i$ mixing angle in matter, $\theta_{_{\!S_i}}^{_{(m)}}$, are defined through:
\begin{equation}\label{NuEffMixAnglei}
\sin2\theta_{_{\!S_i}}^{_{(m)}} = \frac{\sin2\theta_{_{\!S_i}}}{{C}_i^{}} \equiv \frac{\sin2\theta_{_{\!S_i}}}{\;\sqrt{~\big(\!\sin2\theta_{_{\!S_{i\!}}}\big)^2 + \big(\!\cos2\theta_{_{\!S_{i\!}}}\big)^2\left(1-\frac{E_{\nu}}{\,E^\text{res}_{\nu_i}\,}\right)^2\;}}\,.
\end{equation}

\subsubsection{Accelerator constraints on the triple-resonance mechanism}

We begin by discussing how this triple-resonance mechanism addresses shortcoming~\ref{Problem2} of the minimal model. In the high energy limit of $E_\nu\gg E^\text{res}_{\nu_{1,2,3}}$, each active neutrino $\nu_i$ is composed predominantly of the propagation eigenstate ${N}_{i,\, m}^{_{(+)}}$ in~(\ref{NuEigenHi}), with an effective mass-squared given by:
\begin{equation}\label{HighEm2}
m_{i \text{,\,eff\,}}^2 ~=~ 2\,E_\nu\;\omega_i^{_{(+)}} ~\simeq~ m_i^2 + \sin^2\!\theta_{_{\!S_i}}\,\delta M_i^2\left(1+\frac{E^\text{res}_{\nu_i}}{E_\nu}\right)
\end{equation}
From~(\ref{HighEm2}), we see that in the asymptotically high energy limit of $E^\text{res}_{\nu_i}/E_\nu\to 0$, the active mass hierarchy is preserved, i.e., $\delta m_{ij \text{,\,eff\,}}^2\to\delta m_{ij}^2$ ($i,j=1,2,3$), if:
\begin{equation}\label{HighEshift}
\sin^2\!\theta_{_{\!S_1}}\,\delta M_1^2 ~\approx~ \sin^2\!\theta_{_{\!S_2}}\,\delta M_2^2 ~\approx~ \sin^2\!\theta_{_{\!S_3}}\,\delta M_3^2\,.
\end{equation}
While~(\ref{HighEshift}) may appear like an extremely fine-tuned relation, note that if we write explicitly the $\nu_i-S_i$ mixing terms in the Lagrangian,
\begin{equation}
\mathcal{L} \supset \sum_{i = 1,\,2,\,3}~\Big[\;m^{\!\text{\tiny (mix)}\!}_i\,\big(\nu_i\, S_i + \nu_i^c\, S_i^c\big) + M_i\,S_iS_i^c\;\Big] + \cdots\,,
\end{equation}
the mass-mixing term $m^{\!\text{\tiny (mix)}\!}_i$ is essentially $m^{\!\text{\tiny (mix)}\!}_i \approx \sin\!\theta_{_{\!S_i}}M_i$. Hence, when considering this model from a UV completion perspective, one could envision model-building the relations in~(\ref{HighEshift}) by invoking global symmetries, instead of enforcing~(\ref{HighEshift}) via \emph{ad hoc} fine-tuning. These same global symmetries would also provide a natural mechanism for the alignment of $S_i-\nu_i$ mixing with the SM neutrino mass basis.

We can also check the triple-resonance mechanism's prediction for $\delta m_{32 \text{,\,eff\,}}^2$ at T2K in neutrino mode. Following the discussion at the end of subsection~\ref{OscProbSec}, and using~(\ref{HighEm2}) and~(\ref{HighEshift}), we have:
\begin{align}\label{EffM32sqT2Kfull}
\delta m_{32 \text{,\,eff\,}}^2\big|_{_\text{T2K}} &\sim \delta m_{32}^2\big|_{_\text{SM}}\left[\;1 + \frac{\,\sin^2\!\theta_{_{\!S_i}}\,\delta M_i^2\,}{~~~\delta m_{32}^2\big|_{_\text{\tiny SM}}}\, \left(\frac{\, E^\text{res}_{\nu_3}-E^\text{res}_{\nu_2} \,}{\;E_{\nu_\mu}^{\text{\,dip}}\;} \right)\!\Bigg|_{_\text{T2K}}\right]\\[8pt]
 &\sim  \delta m_{32}^2\big|_{_\text{SM}} \times (1\;\pm\;0.08)~~~\text{(from T2K's $\nu$ data),}\nonumber
\end{align}
where the second line in~(\ref{EffM32sqT2Kfull}) is a $2\sigma$-constraint on this prediction to be compatible with T2K's determination of $\delta m_{32}^2$ in neutrino mode~\cite{bibT2K-ml-2021xwb}. Defining
\begin{equation}\label{Rnu}
R_{m_\nu^2} \equiv \frac{\,\sin^2\!\theta_{_{\!S_i}}\,\delta M_i^2\,}{~~~|\delta m_{32}^2|\big|_{_\text{\tiny SM}}}\,,
\end{equation}
remembering that $E_{\nu_\mu}^{\text{\,dip}}\big|_\text{\tiny T2K}\sim 600$\;MeV, and assuming the relation between $E^\text{res}_{\nu_i}\big|_\text{\tiny T2K}$ and $E^\text{res}_{\nu_i}\big|_\text{\tiny MB}$ in~(\ref{EresT2K}), we can recast the constraint in~(\ref{EffM32sqT2Kfull}) as:
\begin{equation}\label{rDeltaEconstraint}
R_{m_\nu^2}\,\left[\frac{~|E^\text{res}_{\nu_3}-E^\text{res}_{\nu_2}|\big|_\text{\tiny MB}}{\text{MeV}}\right]~\lsim~0.8\times10^2\,.
\end{equation}
An analogous calculation for antineutrinos yields
\begin{align}\label{EffM32sqT2KantiNUfull}
\delta \overline{m}_{32 \text{,\,eff\,}}^2\big|_{_{\text{T2K,}\,\bar\nu}} &\sim \delta m_{32}^2\big|_{_\text{SM}}\left[\;1 - \frac{\,\sin^2\!\theta_{_{\!S_i}}\,\delta M_i^2\,}{~~~\delta m_{32}^2\big|_{_\text{\tiny SM}}}\, \left(\frac{\, E^\text{res}_{\nu_3}-E^\text{res}_{\nu_2} \,}{\;E_{\bar\nu_\mu}^{\text{\,dip}}\;} \right)\!\Bigg|_{_\text{T2K}}\right]\\[8pt]
 &\sim  \delta m_{32}^2\big|_{_\text{SM}} \times (1\;\pm\;0.16)~~~\text{(from T2K's $\bar\nu$ data),}\nonumber
\end{align}
which provides a weaker constraint than in~(\ref{EffM32sqT2Kfull}),~(\ref{rDeltaEconstraint}) because T2K's determination of $ \delta m_{32}^2$ in antineutrino mode is less precise~\cite{Berns-ml-2021iss}.

\subsubsection{Reactor constraints on the triple-resonance mechanism}

Independent constraints on the parameters of the triple-resonance mechanism come from lower energy electron antineutrinos at short- and long-baselines reactor-based experiments.

We will not rehash the derivation of the antineutrino propagation eigenstates, which can be easily obtained by reversing the signs of $|\Delta V|$ and $E_{\nu}/E^\text{res}_{\nu_i}$ in~(\ref{subH}),~(\ref{MatterEigeni}), and~(\ref{NuEffMixAnglei}). In the low energy limit of $E_{\bar\nu}\ll E^\text{res}_{\nu_{1,2,3}}$, each active antineutrino $\bar\nu_i$ is composed predominantly of the propagation eigenstate $\bar{N}_{i,\, m}^{_{(-)}}$, with an effective mass-squared given by:
\begin{equation}\label{LowEm2antiNU}
 \overline{m}_{i \text{,\,eff\,}}^2\big|_{{\bar\nu}} ~=~ 2\,E_{\bar\nu}\;\bar\omega_i^{_{(-)}} ~\simeq~ m_i^2 + \sin^2\!\theta_{_{\!S_i}}\delta M_i^2\left(\frac{E_{\bar\nu}}{E^\text{res}_{\nu_i}}\,\cos2\theta_{_{\!S_i}}\!\right)\,.
\end{equation}

With~(\ref{LowEm2antiNU}), we can revisit the prediction for $\delta m_{31}^2$ from short-baseline reactor measurements (by, e.g., Daya Bay and RENO), discussed at the end of subsection~\ref{ToyAntiNU}. Using Daya Bay for concreteness, we have,
\begin{align}\label{EffM31sqReactorFull}
\delta \overline{m}_{31 \text{,\,eff\,}}^2\big|_{\text{reactor }{\bar\nu}_e} &\simeq \delta m_{31}^2\big|_{_\text{SM}}\,\left[\;1 + R_{m_\nu^2}\;\frac{|\delta m_{32}^2|}{\,\delta m_{31}^2}\Bigg|_{_\text{SM}}\!\left(\!\frac{E_{\bar\nu_e}^{\text{\,dip}}}{\;E^\text{res}_{\nu_3}\;}-\frac{E_{\bar\nu_e}^{\text{\,dip}}}{\;E^\text{res}_{\nu_1}\;}\!\right)\!\Bigg|_{_\text{DayaB}}\right]\qquad\\[8pt]
 &\sim  \delta m_{31}^2\big|_{_\text{SM}} \times (1\;\pm\;0.05)~~~\text{(from Daya Bay's $\bar\nu_e$ data at $2\sigma$).}\nonumber
\end{align}
Above, we made the approximation $\cos2\theta_{_{\!S_i}}\approx 1$ and used~(\ref{Rnu}).
Using $E_{\bar\nu_e}^{\text{\,dip}}\big|_\text{\tiny DayaB}\sim3$\;MeV, we can recast the constraint in the second line of~(\ref{EffM31sqReactorFull}) as:
\begin{equation}\label{rDeltaEDayaBay}
R_{m_\nu^2}\,\left[\frac{~|E^\text{res}_{\nu_3}-E^\text{res}_{\nu_1}|\big|_\text{\tiny MB}}{\text{MeV}}\right]~\lsim~1.5\times10^3\,,
\end{equation}
which is about an order-of-magnitude weaker than~(\ref{rDeltaEconstraint}).

As for reactor $\bar\nu_e$'s at long-baselines, of interest to us is KamLAND's determination of the `solar' mass-squared splitting $\delta m_{21}^2$ with an error of $\sim$2.5\%~\cite{KamLAND-ml-2013rgu}. Using~(\ref{LowEm2antiNU}), we can obtain the triple-resonance mechanism's prediction for $\delta \overline{m}_{21 \text{,\,eff\,}}^2$ at KamLAND:
\begin{align}\label{EffM21sqantiNUfull}
\delta \overline{m}_{21 \text{,\,eff\,}}^2\big|_{_{\text{KamLAND}}} &\simeq \delta m_{21}^2\big|_{_\text{SM}}\left[\;1 + R_{m_\nu^2}\;\frac{|\delta m_{32}^2|}{\,\delta m_{21}^2}\Bigg|_{_\text{SM}}\!\left(\!\frac{E_{\bar\nu_e}}{\;E^\text{res}_{\nu_2}\;}-\frac{E_{\bar\nu_e}}{\;E^\text{res}_{\nu_1}\;}\!\right)\!\Bigg|_{_\text{KamL}}\right]\qquad\\[8pt]
 &\sim  \delta m_{21}^2\big|_{_\text{SM}} \times (1\;\pm\;0.05)~~~\text{(from KamLAND's $\bar\nu_e$ data at $2\sigma$).}\nonumber
\end{align}
KamLAND observes $\bar\nu_e$'s with energies ranging from $E_{\bar\nu_e}\sim(0.9-7)$\;MeV\@. As a simplification, we will use the energy of maximal $\bar\nu_e$ disappearance at KamLAND as an input in~(\ref{EffM21sqantiNUfull}), namely, $E_{\bar\nu_e}^{\text{\,dip}}\big|_\text{\tiny KamL}\sim3.6$\;MeV\@. Making the additional assumption that the resonant energies $E^\text{res}_{\nu_i}$ for KamLAND and T2K are the same, and using~(\ref{EresT2K}), we can recast the constraint in~(\ref{EffM21sqantiNUfull}) as:
\begin{equation}\label{rDeltaEkamLAND}
R_{m_\nu^2}\,\left[\frac{~|E^\text{res}_{\nu_2}-E^\text{res}_{\nu_1}|\big|_\text{\tiny MB}}{\text{MeV}}\right]~\lsim~23\,.
\end{equation}

\subsection{Full model --- part II: the `vanilla' off-resonance component}
\label{VanillaPart}

We now complete the description of the full model by introducing its last component, namely, a sterile neutrino that does not couple to dark vector mediators and therefore does not experience the matter potential in~(\ref{Vm}). We will denote this sterile species as $S_0$, and will often refer to it as the \emph{vanilla} sterile neutrino.

In a generic UV realization of the off-resonance mechanism, $S_0$ will mix with a linear combination of the active neutrinos, $\nu_1$, $\nu_2$, and $\nu_3$, with a mixing angle $\theta_{_{\!S_0}}$. Its purpose is to induce non-resonant oscillations that could explain the $\bar\nu_e$-like event excess at LSND and MiniBooNE, as well as the mid-energy $\nu_e$-like event excess at MiniBooNE, through $\nunubar_{\!\mu}\to \SSbar_{\!0}\to\nunubar_{\!e}$ oscillations. This vanilla off-resonance mechanism will also lead to predictions for $\nunubar_{\!\mu}$ and $\nunubar_{\!e}$ disappearance in a variety of other experiments, such as MINOS, IceCube, BEST, RENO/NEOS, etc.\ Since the resonant and non-resonant mechanisms do not interfere in any significant way, one could in principle fit the parameters of the vanilla sterile neutrino independently of the resonant part of the model. We will not perform this exercise in this paper, however, since our focus is on the novel part of the resonance mechanism.

Instead, we will fix the parameters of the off-resonance component to a simplified benchmark that will allow us to solve for all the propagation eigenstates \emph{analytically}. In this simplified benchmark, the active state mixing with $S_0$ will consist exclusively of $\nu_1$. Consider initially the limit in which we turn off the mixing of $\nu_1$ with $S_1$ (i.e., $\theta_{_{\!S_1}}\! \to 0$). The two mass eigenstates would then be:
\begin{subequations}\label{nu1primenuS}
\begin{alignat}{3}
|\nu^\prime_{1\,}\rangle &= \cos\theta_{_{\!S_0}}\,|\nu_1\rangle  -  \sin\theta_{_{\!S_0}}\,|S_0\rangle\,,\qquad &{\omega}^\prime_{1}&~~\xrightarrow[\,\theta_{_{\!S_1}}\! \to 0\;]{}~~\frac{m_1^2}{\,2E_{\nu}\,} \label{nu1prime}\\
|\nu_{S\,}\rangle &= \sin\theta_{_{\!S_0}}\,|\nu_1\rangle  +  \cos\theta_{_{\!S_0}}\,|S_0\rangle\,,\qquad &{\omega}_0 &~~\xrightarrow[\,\theta_{_{\!S_1}}\! \to 0\;]{}~~\frac{M_0^2}{\,2E_{\nu}\,}\label{nuS}
\end{alignat}
\end{subequations}
The choice of $\nu_1-S_0$ mixing is motivated by a combination of anomalies and experimental bounds. First, since $\nu_1$ has a significant $\nu_e$-component, $\nu_1-S_0$ mixing can easily accommodate a large (off-resonance) $\nu_e$ disappearance rate,  $\overline{P}(\nu_e\not\to\nu_e) \approx  \overline{P}(\nu_e\to S_0) =  1/2\times|U_{e 1}|^2 \sin^22\theta_{_{\!S_0}}$, which is convenient to address the Gallium anomalies, for instance~\cite{GALLEX-ml-1997lja,SAGE-ml-1998fvr,Abdurashitov-ml-2005tb,Kaether-ml-2010ag,Giunti-ml-2012tn,Barinov-ml-2021mjj}. Secondly, $\nu_1$ has a smaller $\nu_\mu$-component than $\nu_2$ and $\nu_3$, and therefore $\nu_1-S_0$ mixing is subject to weaker $\nunubar_{\!\mu}$ disappearance constraints from, e.g., MINOS~\cite{MINOS-ml-2017cae} and IceCube~\cite{IceCube-ml-2016rnb,IceCube-ml-2020tka}.

\begin{figure}
\centering
\includegraphics[width=1.0\textwidth]{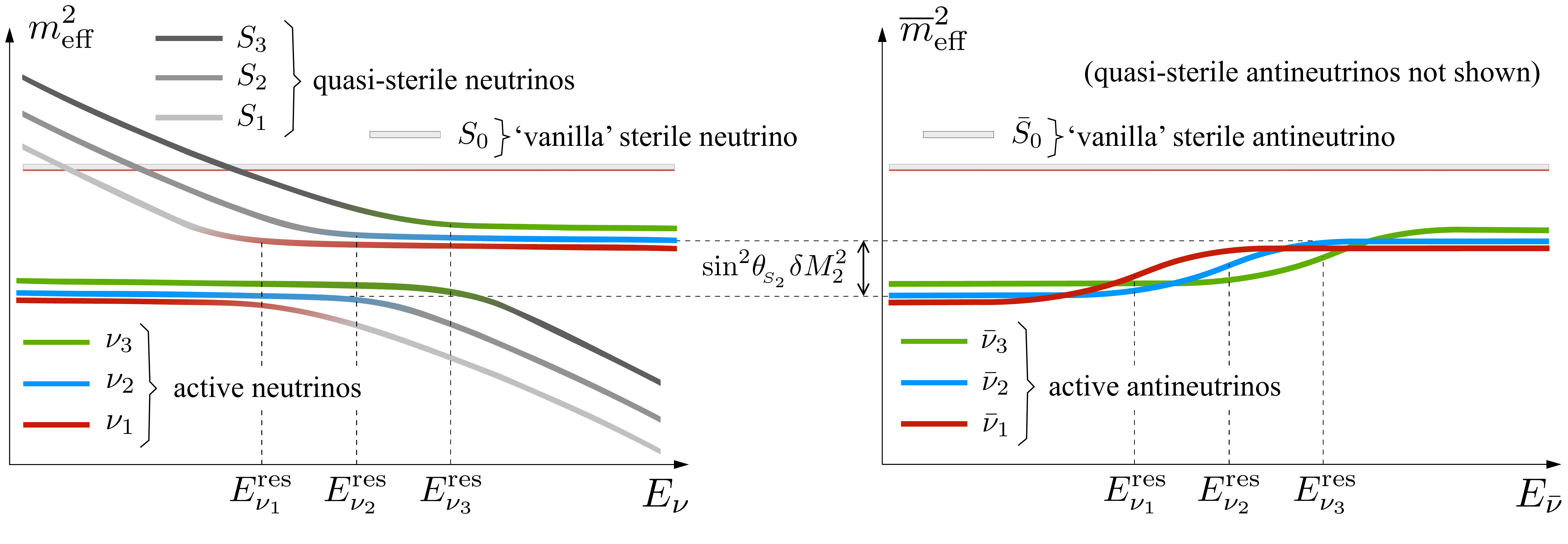}
\caption{[\emph {For illustration purposes only --- not to scale\,}] Effective mass-squared eigenvalues of neutrino propagation eigenstates (left panel) and antineutrino propagation eigenstates (right panel) as a function of (anti-)neutrino energy in the full model. The eigenstates' flavor compositions are denoted by their color shading.\label{CartoonFullModel}}
\end{figure}

Given~(\ref{nu1primenuS}), we could now straightforwardly turn on $\nu_1-S_1$ mixing. However, this would then require numerically diagonalizing the Hamiltonian for the $\{\nu_1,S_0,S_1\}$ subsystem in order to obtain the propagation eigenstates. In order to avoid loosing analytical control over our treatment, we will introduce a second simplification to our benchmark: we will mix $S_1$ with $\nu^\prime_1$ in~(\ref{nu1prime}) instead. This will allow us to retain our analytical expressions for the propagation eigenstates at arbitrary matter densities by simply making the replacement $\nu_1\to\nu^\prime_1$ in~(\ref{fullH}),~(\ref{subH}), and~(\ref{MatterEigeni}). This small deformation will not qualitatively affect any of the important features of the resonance mechanism. Quantitatively, we have checked that if $S_1$ were to mix with $\nu_1$ instead of $\nu_1^\prime$, this would shift best fit parameters by $\OO(10\%)$, and would also require an $\OO(10\%)$ adjustment of relation~(\ref{HighEshift}) in order to preserve the in-medium active neutrino mass hierarchy at high energies.

Finally, we briefly return to the issue of potential CP violating phases introduced by mixings of the active neutrinos with the sterile and quasi-sterile states. In analogy with the toy model case, complex phases $\delta_{_{\!S_i}}$ from $\nu_i-S_i$ mixing are unphysical and can be absorbed by unitary rotations of the $|S_i\rangle$ states. This is possible because, by construction, $\nu_i-S_i$ mixing is aligned with the active Yukawa basis, i.e., there is no ``cross-mixing'' of the type $\nu_j-S_k$ with $j\neq k$. Similarly, the unphysical complex phase from $\nu_1-S_0$ mixing in~(\ref{nu1primenuS}) can be absorbed by a unitary rotation of $|S_0\rangle$. Therefore, in the benchmark models we will be considering in the next section, no additional sources of CP-violation exist beyond the SM CP-violating phase $\delta_{\text{CP}}$. However, more generally, a more anarchical mixing pattern between $S_0$ and a linear combination of $\nu_i$ and $S_i$ states will introduce additional physical CP-violating phases. The complex phases from $S_0-S_i$ mixing, in particular, will only be relevant in the very narrow resonance window, and while they may introduce quantitative modifications to the resonant oscillation pattern, they will not qualitatively affect the overall mechanism's ability to address the MiniBooNE excess. Outside the resonance window, the complex phases from $S_0-S_i$ mixing will have a negligible effect in active neutrino appearance and disappearance probabilities due to the suppressed $S_0-S_i$ and $\nu_i-S_i$ vacuum mixing angles. In contrast, the complex phases from $S_0-\nu_i$ mixing will be relevant outside the resonance window, where neutrino oscillations are well described by the standard `3+1' model. Such effects have been considered elsewhere in the literature (see, e.g.,~\cite{Donini-ml-1999jc,Klop-ml-2014ima,Cianci-ml-2017okw,Palazzo-ml-2020apc}), and are not the focus of the present study.

Figure~\ref{CartoonFullModel} illustrates the components of the full model, including the neutrino and antineutrino propagation eigenstates, their eigenvalues, and their flavor composition. The relation~(\ref{HighEshift}), combined with the fact that all active neutrinos cross the resonance within a narrow energy range, results in a much smaller deviation of the in-medium active mass spectrum relative to the SM expectation. Also, note that unlike the toy model discussed in section~\ref{toymodel}, the in-medium mass hierarchy in this full model realization does not `flip' at high~energies.

\subsection{Full model --- part III: fitting the MiniBooNE low energy excess}\label{SecMBfit}

With the full model ingredients introduced, we can recast the active neutrino flavor eigenstates in terms of the propagation eigenstates in order to derive $\nu_\mu\to\nu_e$ oscillation probabilities for MiniBooNE\@.
With $\ell$ denoting active neutrino flavors ($\ell=e,\mu,\tau$), we have:
\begin{eqnarray}\label{nuFullL}
|\nu_\ell(L)\rangle~=~~~\sin\theta_{_{\!S_0}}\!\!\!&&U^*_{\ell 1\,}\,\Big[\;e^{-i\,\omega_0 \,(L\,+\,\overline{L}_\text{\tiny decay})}\;|\nu_{S\,}\rangle~\Big] \\[4pt]
~~\,+\,\cos\theta_{_{\!S_0}}\!\!\!&&U^*_{\ell 1\,}\, \Big[\,\cos\theta_{_{\!S_1}}^{_{(m)}}\,e^{-i\,\omega_1^{_{(-)}} L}\;|N_{1,\, m}^{_{(-)}}\rangle ~+~ \sin\theta_{_{\!S_1}}^{_{(m)}}\,e^{-i\,\omega_1^{_{(+)}} L}\;|N_{1,\, m}^{_{(+)}}\rangle~\Big] \qquad \nonumber\\[4pt]
+&&U^*_{\ell 2\,}\, \Big[\,\cos\theta_{_{\!S_2}}^{_{(m)}}\,e^{-i\,\omega_2^{_{(-)}} L}\;|N_{2,\, m}^{_{(-)}}\rangle ~+~ \sin\theta_{_{\!S_2}}^{_{(m)}}\,e^{-i\,\omega_2^{_{(+)}} L}\;|N_{2,\, m}^{_{(+)}}\rangle~\Big] \qquad \nonumber\\[4pt]
+&&U^*_{\ell 3\,}\, \Big[\,\cos\theta_{_{\!S_3}}^{_{(m)}}\,e^{-i\,\omega_3^{_{(-)}} L}\;|N_{3,\, m}^{_{(-)}}\rangle ~+~ \sin\theta_{_{\!S_3}}^{_{(m)}}\,e^{-i\,\omega_3^{_{(+)}} L}\;|N_{3,\, m}^{_{(+)}}\rangle~\Big]\,. \qquad \nonumber
\end{eqnarray}
Above, the matter mixing angles $\theta_{_{\!S_i}}^{_{(m)}}$ are defined in~(\ref{NuEffMixAnglei}), and the propagation eigenphases $\omega_i^{_{(\pm)}}$ and $\omega_0$ are defined in~(\ref{MatterEigeni}) and~(\ref{nuS}), respectively. As before, we take MiniBooNE's propagation length $L$ to be the average distance through which neutrinos travel in matter, $L\approx 490\,$m, and we assume constant matter density and earth composition throughout.\footnote{In particular, we neglect possible effects of other structures in the neutrino's propagation path, such as the steel and concrete beam absorber, vault and detector walls, the mineral oil target before interaction,~etc.} The additional `averaged decay baseline' $\overline{L}_\text{\tiny decay}$ appearing in the first line of~(\ref{nuFullL}) is an approximation to account for the state's evolution within the decay region of the BNB beamline, which for simplicity we take to be $\overline{L}_\text{\tiny decay} \sim 30\,$m. Only the phase of the $\nu_S$ component of $\nu_\ell$ changes non-negligibly within the decay region (see~(\ref{nu1primenuS})); therefore, the phase change of the $\nu_1^\prime$ component within the decay region is ignored in~(\ref{nuFullL}).

Given the relations~(\ref{tripleResDeltaV}),~(\ref{HighEshift}), and~(\ref{Rnu}), we can fully specify all the parameters of the full model by a set of 7 independent parameters: 5 related to the triple-resonance mechanism, which we choose to be $\theta_{_{\!S_2}}$, $R_{m_\nu^2}$, $E^\text{res}_{\nu_1}$, $E^\text{res}_{\nu_2}$, $E^\text{res}_{\nu_3}$; and 2 related to the off-resonance oscillations, $\theta_{_{\!S_0}}$ and $M^2_0$.

In fitting MiniBooNE's $\nu_e$ appearance data, we will fix the off-resonance parameters to the following (non-optimized) values:\footnote{This choice of off-resonance parameters is motivated by the Gallium $\nu_e$ disappearance anomalies~\cite{GALLEX-ml-1997lja,SAGE-ml-1998fvr,Abdurashitov-ml-2005tb,Kaether-ml-2010ag,Giunti-ml-2012tn,Barinov-ml-2021mjj}, including the most recent results from the BEST collaboration~\cite{Barinov-ml-2021asz}. However, we note that these parameters are in strong tension with more recent oscillation bounds from reactor-based antineutrino experiments~\cite{Giunti-ml-2021kab}, such as the combined results from RENO and NEOS~\cite{RENO-ml-2020hva}. Notwithstanding, the $\nu_e$ spectral shape at RENO, NEOS, and other reactor experiments (Daya Bay, Double Chooz) shows a significant departure from flux models' expectations at energies around $\sim5-6\;$MeV, whose origin is not understood~\cite{NEOS-ml-2016wee,RENO-ml-2016ujo,DayaBay-ml-2016ssb,Huber-ml-2016xis,Zacek-ml-2018bij,Qian-ml-2018wid,DoubleChooz-ml-2019qbj}. We speculate that there might be additional systematic uncertainties in reactor antineutrino measurements that have not been accounted for~\cite{Hayes-ml-2013wra,Dwyer-ml-2014eka,Novella-ml-2015eaw,Hayes-ml-2016qnu}; therefore, we choose to ignore reactor-based constraints on `3+1' sterile neutrino models~\cite{Dentler-ml-2017tkw,Giunti-ml-2017yid,Giunti-ml-2020uhv,Serebrov-ml-2021ndf}.}
\begin{equation}\label{LSNDparam}
\theta_{_{\!S_0}} = 0.30\,, \qquad M^2_0 = 1.0\;\text{eV}^2\,.
\end{equation}

\begin{figure}
\includegraphics[width=1\textwidth,trim= 0 0 70 0,clip]{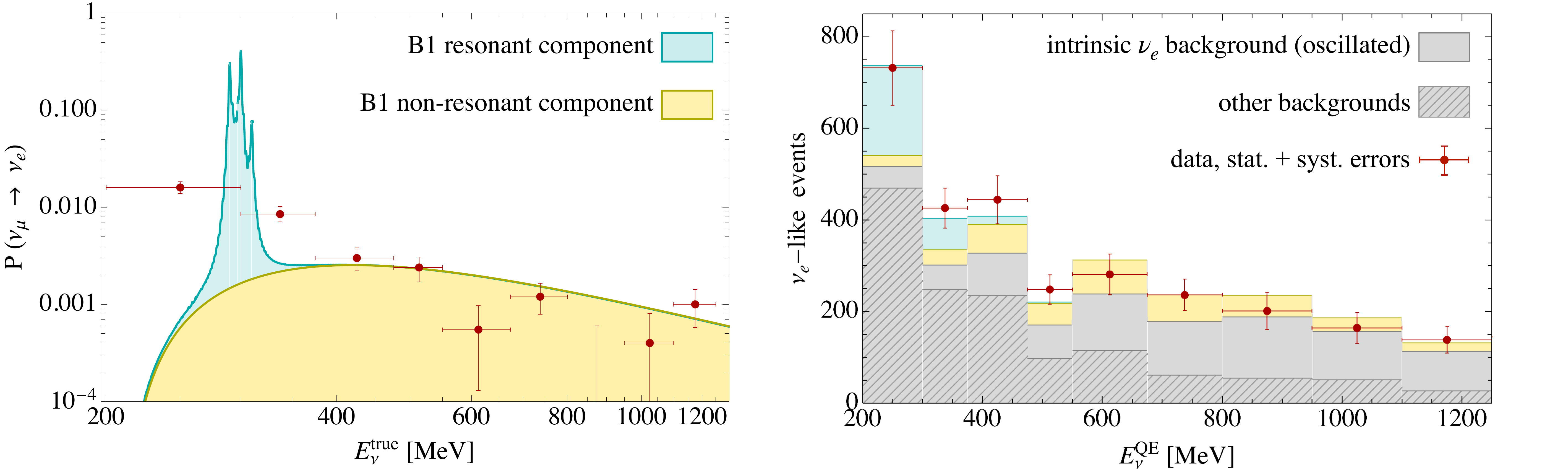}
\caption{Left panel: the $\nu_\mu\to\nu_e$ appearance probability at MiniBooNE predicted by the benchmark model B1 specified in~(\ref{LSNDparam}),~(\ref{B1}). The non-resonant component (yellow region) is the expectation from this model if the triple-resonance mechanism were `turned off' (i.e., if the active-quasi-sterile mixing angles $\theta_{_{\!S_1}},\theta_{_{\!S_2}},\theta_{_{\!S_3}}$ were set to zero). The red data points with error bars were taken from figure\,21 of~\cite{MiniBooNE-ml-2020pnu}; they assume $E_\nu^\text{true}\sim E_\nu^\text{\tiny QE}$, which is a very coarse approximation. As such, they should be interpreted as a \emph{qualitative} measure of MiniBooNE's anomalous appearance probabilities. Right panel: the expected $\nu_e$ appearance event rate at MiniBooNE predicted by the full model benchmark B1. The intrinsic (beam) $\nu_e$ component of the background takes $\nu_e\not\to\nu_e$ oscillations into account. The red data points correspond to the measurements reported by MiniBooNE in~\cite{MiniBooNE-ml-2020pnu}, and the error bars include both statistical and systematic uncertainties.
\label{B1MBfit}}
\end{figure}

\begin{figure}
\includegraphics[width=1\textwidth,trim= 0 0 70 0,clip]{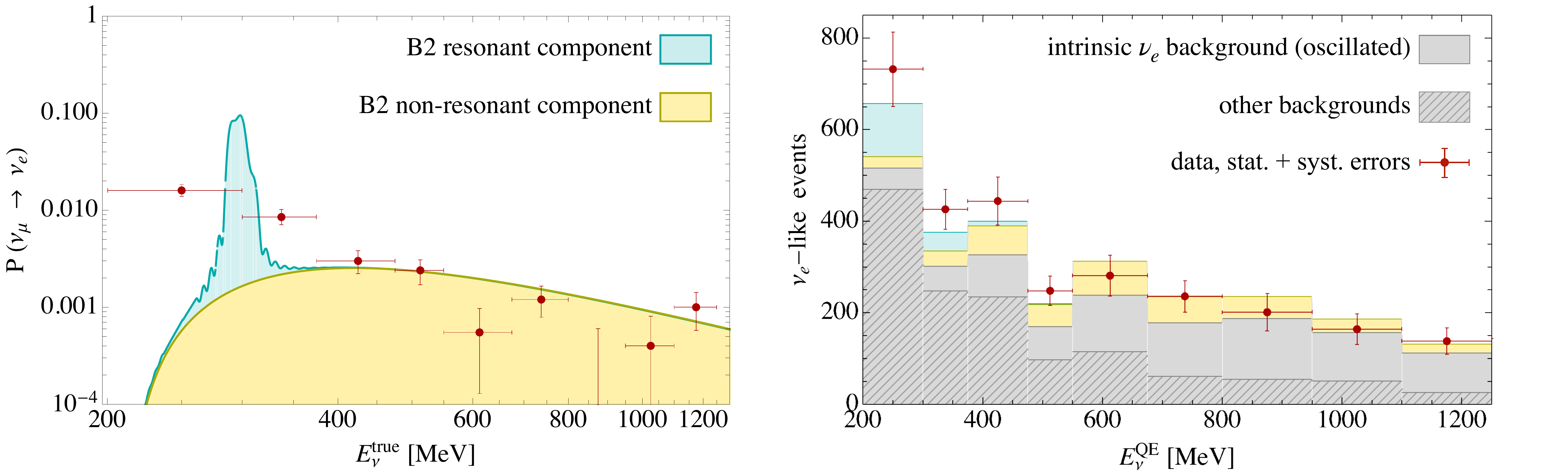}
\caption{Same as in figure~\ref{B1MBfit} but for the benchmark model B2 specified in~(\ref{LSNDparam}),~(\ref{B2}).\label{B2MBfit}}
\end{figure}

As for the parameters of the triple-resonance mechanism, we have found that if either $E^\text{res}_{\nu_1}$ or $E^\text{res}_{\nu_2}$ is within $\mathcal{O}(20\%)$ of $ \sim $300\;MeV, the full model can provide excellent fits to MiniBooNE's data while simultaneously satisfying the constraints in~(\ref{rDeltaEconstraint}),~(\ref{rDeltaEDayaBay}), and~(\ref{rDeltaEkamLAND}). These excellent fits (i.e., within $1\sigma$ of the best-fit point) are achieved if the mixing angle $\theta_{_{\!S_j}}$ associated with the resonant energy $E^\text{res}_{\nu_j}$ closest to $ \sim $300\;MeV satisfies the condition of appearance probability saturation at the resonance, analogous to~(\ref{Theta3max}):
\begin{align}\label{Thetaimax}
\sin\theta_{_{\!S_j}}&~\lesssim~\big|{\sin\theta_{_{\!S_j}}^{\text{(max)}}}\big| ~=~ \frac{\,R_{m_\nu^2}\,|\delta m_{32}^2|\,}{2\,\pi}~\frac{L}{\,E^\text{res}_{\nu_j}\,}\Bigg|_\text{MB}\\
&~\lesssim~ 3.3\times 10^{-3}\;R_{m_\nu^2}\;\frac{\,(300\;\text{MeV})\,}{~~~~E^\text{res}_{\nu_j}\big|_{_{\text{MB}}}\,}\,.\nonumber
\end{align}
Figures~\ref{B1MBfit} and~\ref{B2MBfit} (left panels) show the $\nu_\mu\to\nu_e$ appearance probability at MiniBooNE as a function of true neutrino energy for two illustrative benchmark points:
\begin{subequations}\label{Benchmarks}
\begin{alignat}{6}
\text{B1}&:\quad& \theta_{_{\!S_2}}&={\bf{0.003}},&\quad R_{m_\nu^2}&=0.9,&\quad E^\text{res}_{\nu_1}&=290\;\text{MeV},&\quad E^\text{res}_{\nu_2}&=300\;\text{MeV},&\quad E^\text{res}_{\nu_3}&=310\;\text{MeV},\label{B1}\\
\text{B2}&:\quad& \theta_{_{\!S_2}}&={\bf{0.006}},&\quad R_{m_\nu^2}&=0.9,&\quad E^\text{res}_{\nu_1}&=290\;\text{MeV},&\quad E^\text{res}_{\nu_2}&=300\;\text{MeV},&\quad E^\text{res}_{\nu_3}&=310\;\text{MeV},\label{B2}
\end{alignat}
\end{subequations}
with the off-resonance parameters set to the values in~(\ref{LSNDparam}). The right panels in figures~\ref{B1MBfit} and~\ref{B2MBfit} show the expected contribution of these two benchmarks to the MiniBooNE $\nu_e$ event spectrum. These rates were obtained using the information provided by the MiniBooNE collaboration in its 2021 public data release~\cite{hepdata.114365.v1}, and include the effects of oscillations of the intrinsic $\nu_e$ background from the Booster Neutrino Beam (BNB)\@. Note that even for benchmark B2 in~(\ref{B2}), which lies outside the appearance probability saturation range in~(\ref{Thetaimax}), the triple-resonance mechanism provides an adequate fit to the MiniBooNE $\nu_e$-like excess, as can be seen from figure~\ref{B2MBfit} (left panel).

At this point, it is germane to comment on the $\cos\theta_e$ distribution of MiniBooNE's excess events ($\theta_e$ is defined as the angle of the reconstructed electron relative to the neutrino beam). The $\cos\theta_e$ spectral shape of the MiniBooNE excess appears to be more forward-peaked in the region $\cos\theta_e>0.9$ than what is expected from the kinematics of $\nu_e$ charged-current quasi-elastic (CCQE) scattering, but not forward enough to be consistent with $\nu-e^-$ elastic scattering. In fact, the forward excess in $\cos\theta_e>0.9$ has motivated many non-oscillatory interpretations of the MiniBooNE anomaly. Indeed, if statistically significant, this spectral feature could in principle rule out explanations of the $\nu_e$-like excess based on \emph{bona fide} $\nu_e$ appearance. However, we note that the MiniBooNE collaboration has not performed any studies to quantify the experimental systematics associated with the $\cos\theta_e$ distribution (and, in fact, MiniBooNE's modeling of the $\nu_e$ backgrounds in the forward low energy kinematic region has limited statistics). Hence, at present one cannot rule out the $\nu_e$ CCQE scattering hypothesis for the MiniBooNE excess based on its $\cos\theta_e$ event distribution.

\begin{figure}
\centering
\includegraphics[width=0.75\textwidth,trim= 0 0 50 0,clip]{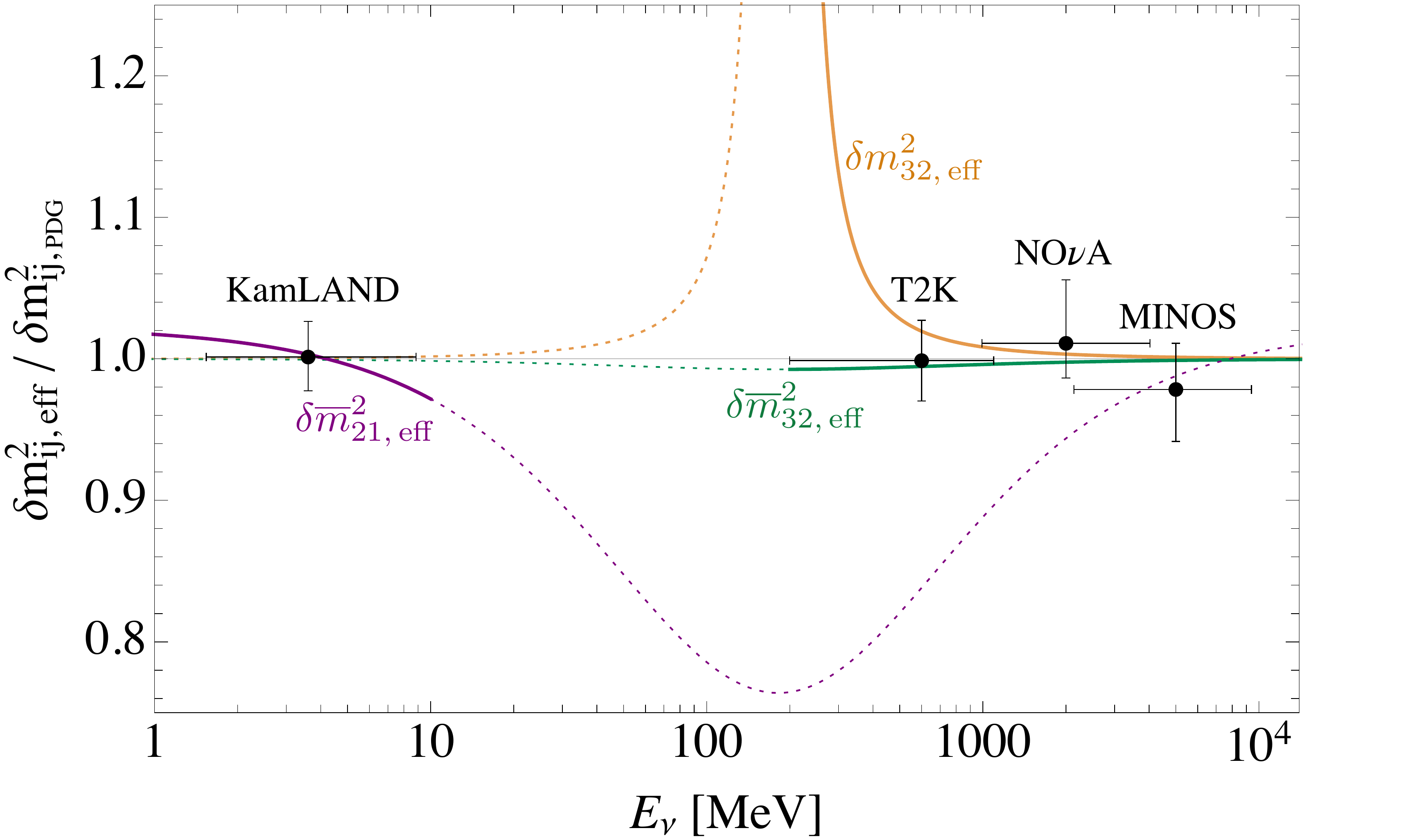}
\caption{Effective mass-squared splittings $\delta m_{ij \text{,\,eff}}^2$ predicted by the full model benchmarks B1 and B2 in~(\ref{Benchmarks}) as a function of (anti-)neutrino energy (both benchmarks' predictions are indistinguishable as displayed in this plot). All effective mass-squared splittings shown have been normalized to the appropriate central value $\delta m_{ij, \text{\tiny \,PDG}}^2$ reported in the PDG~\cite{ParticleDataGroup-ml-2020ssz}. Solid purple curve: the effective `solar' mass-squared splitting for antineutrinos, $\delta \overline{m}_{21 \text{,\,eff}}^2$, expected at KamLAND\@. Solid orange (green) curve:  the effective `atmospheric' mass-squared splitting for neutrinos (antineutrinos), $\delta m_{32 \text{,\,eff}}^2$ ($\delta \overline{m}_{32 \text{,\,eff}}^2$), expected at long-baseline accelerator neutrino experiments running in neutrino (antineutrino) mode. The dotted parts of these curves indicate energy ranges that have not been experimentally probed. The data points with error bars denote recent measurements of $\delta \overline{m}_{21 \text{,\,eff}}^2$, $\delta m_{32 \text{,\,eff}}^2$, and $\delta \overline{m}_{32 \text{,\,eff}}^2$ by long-baseline experiments~\cite{KamLAND-ml-2013rgu,bibT2K-ml-2018rhz,MINOS-ml-2020llm,NOvA-ml-2021nfi}.\label{FullMeff}}
\end{figure}

Finally, we look at how neutrino oscillations in the full model can be satisfactorily reconciled with data from long-baseline experiments. In figure~\ref{FullMeff} we show the effective active mass-squared splittings $\delta m_{32 \text{,\,eff\,}}^2$, $\delta \overline{m}_{32 \text{,\,eff\,}}^2$, and $\delta \overline{m}_{21 \text{,\,eff\,}}^2$ expected in long-baseline experiments as a function of (anti-)neutrino energy for the benchmarks in~(\ref{Benchmarks}). The rapid relaxation of these active mass-squared splittings towards their SM expectation indicates that the triple-resonance mechanism, unlike the models discussed in~\cite{Karagiorgi-ml-2012kw,Asaadi-ml-2017bhx,Doring-ml-2018cob,Barenboim-ml-2019hso}, can explain the MiniBooNE excess while remaining compatible with observations from long-baseline neutrino experiments.

\begin{figure}
\centering
\includegraphics[width=0.75\textwidth,trim= 0 0 50 0,clip]{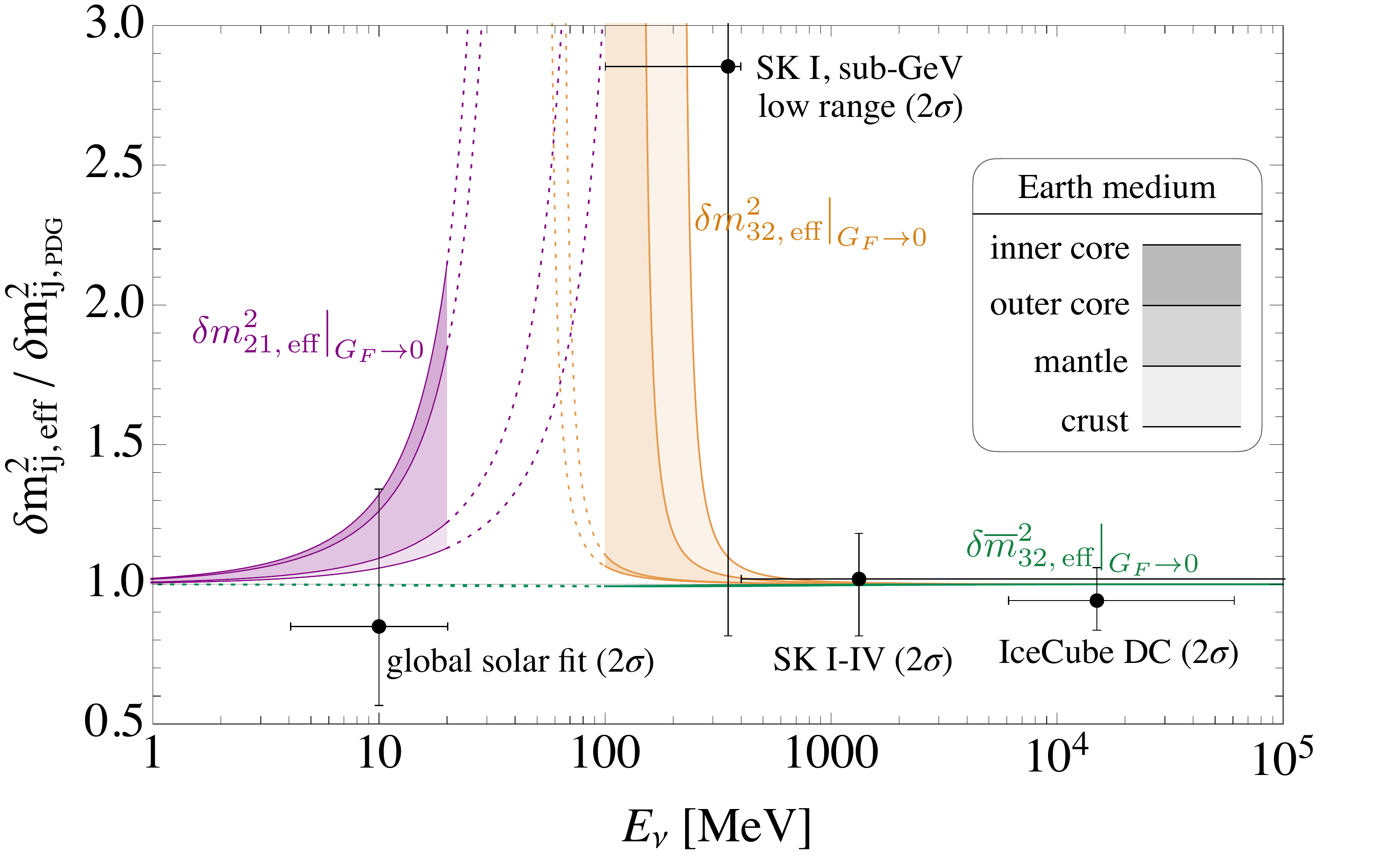}
\caption{Similar to figure~\ref{FullMeff}, but for atmospheric (anti-)neutrinos and solar neutrinos propagating through different media within the Earth, and ignoring the SM matter effects (i.e., setting $G_F=0$). The left-most data point corresponds to the global fit to solar neutrino data, including day-night asymmetries, obtained in~\cite{Esteban-ml-2020cvm}. The other data points correspond to atmospheric neutrino measurements from Super-Kamiokande~\cite{Super-Kamiokande-ml-2005mbp,Super-Kamiokande-ml-2017yvm} and IceCube DeepCore~\cite{IceCube-ml-2017lak}.\label{MeffAtmSolar}}
\end{figure}

In figure~\ref{MeffAtmSolar} we show the effective active mass-squared splittings $\delta m_{32 \text{,\,eff\,}}^2$ ($\delta \overline{m}_{32 \text{,\,eff\,}}^2$) for atmospheric (anti-)neutrinos and $\delta {m}_{21 \text{,\,eff\,}}^2$ for solar neutrinos when propagating through different media within the Earth. These effective mass-squared splittings are obtained in the limit $G_F\to 0$, and should replace the corresponding vacuum values in the calculation of terrestrial MSW matter effects. Specifically, we show the effective mass-squared splittings $\delta m_{ij \text{,\,eff\,}}^2\big|_{G_F\to 0}$ in the Earth's inner core ($\rho=13.0\,\text{g/cm}^3$), outer core ($\rho=11.3\,\text{g/cm}^3$), mantle ($\rho=5.0\,\text{g/cm}^3$), and crust ($\rho=3.3\,\text{g/cm}^3$). The assumed densities for these various media follow the PREM model~\cite{Dziewonski} used in the atmospheric and solar neutrino analyses by the Super-Kamiokande collaboration. Figure~\ref{MeffAtmSolar} also shows the global solar neutrino data determination of $\delta m_{21}^2$~\cite{Esteban-ml-2020cvm}, and the determination of $\delta m_{32}^2$ by Super-Kamiokande runs I--IV~\cite{Super-Kamiokande-ml-2017yvm} and by IceCube DeepCore~\cite{IceCube-ml-2017lak}. Note that despite the deformation of the active neutrino spectrum near the triple-resonance region, the effective active mass-squared splittings $\delta m_{ij \text{,\,eff\,}}^2\big|_{G_F\to 0}$ remain compatible with experimental determinations of $\delta m_{ij}^2$ within the energy ranges probed. In particular, deviations from the SM expectation are strongly suppressed for atmospheric antineutrinos, as well as for atmospheric neutrinos in the mid-to-high energy range $E_\nu\gtrsim 400$\,MeV\@. For sub-GeV atmospheric neutrinos in the low energy range $E_\nu\lesssim 200$\,MeV, the active mass spectrum departs significantly from the SM expectation; however, in this energy range $\delta m_{32}^2$ is very poorly constrained by the zenith-angle dependence of the atmospheric neutrino flux~\cite{Super-Kamiokande-ml-2005mbp}. In the much lower energy regime relevant for solar neutrinos, the range of $\delta {m}_{21 \text{,\,eff\,}}^2\big|_{G_F\to 0}$ in the Earth's media (inner/outer core, mantle, and crust) is in moderate tension ($\lesssim 2\sigma$) with the global solar neutrino data; this can be traced back to the slight discrepancy in the favored range for $\delta m_{21}^2$ between solar data \emph{vs} KamLAND~\cite{KamLAND-ml-2013rgu}. Note, however, that the global solar fit to $\delta m_{21}^2$ assumes that the flux of neutrinos emitted by the Sun is the one predicted in the SM\@. This flux might be modified due to the new matter effects induced in the solar medium by the dark sector vector mediator; in fact, some UV completions of the dark sector matter potential could alleviate this mild tension in the determination of $\delta m_{21}^2$. This topic will be considered in further detail in future work~\cite{future}.

\section{Further implications of quasi-sterile neutrinos and BSM matter effects}
\label{SunMicroBooNE}

There are other experimental observations that could be significantly impacted by the resonant mechanism of quasi-sterile neutrinos, and in particular by the specific realization considered in section~\ref{FullModel} to explain the MiniBooNE and LSND anomalies. Two obvious issues are (i) the implications for solar neutrinos, and (ii) the expected BSM $\nu_e$ flux at MicroBooNE, which is of significant relevance at present  in light of the recent results reported by the MicroBooNE collaboration in~\cite{MicroBooNE-ml-2021rmx}. In this section we briefly comment on generic experimental implications of the triple-resonance mechanism for solar neutrinos and MicroBooNE, while deferring a comprehensive investigation of these issues to a future~publication~\cite{future}.

\subsection{Implications for MicroBooNE}\label{MicroBooNE}

The MicroBooNE collaboration recently released the results of three searches for $\nu_e$ scattering events based on data obtained with Fermilab's BNB in neutrino-mode from an exposure of $7\times10^{20}$ POT~\cite{MicroBooNE-ml-2021nxr,MicroBooNE-ml-2021jwr,MicroBooNE-ml-2021sne}. These analyses targeted independent final states of $\nu_e$-nuclear scattering: the `Wire-Cell' search in~\cite{MicroBooNE-ml-2021nxr} was fully inclusive in all possible hadronic final states; the `Deep-Learning' search in~\cite{MicroBooNE-ml-2021jwr} selected exclusive two-body $e^-p$ final states to obtain increased sensitivity to $\nu_e$ CCQE scattering; and the semi-inclusive `Pandora' search in~\cite{MicroBooNE-ml-2021sne} applied a pion veto on its selection of $e^-+Np$ final states ($N\geq0$). The non-observation of an excess of events in these searches (and, in fact, the observation of a \emph{deficit} of events in the inclusive Wire-Cell and CCQE Deep-Learning searches) allowed MicroBooNE to exclude an \emph{ad hoc} parametrization of the MiniBooNE low energy excess, coined the `empirical LEE' model, at a confidence level greater than $\sim97\%$.

\begin{figure}
\centering
\includegraphics[width=1.0\textwidth]{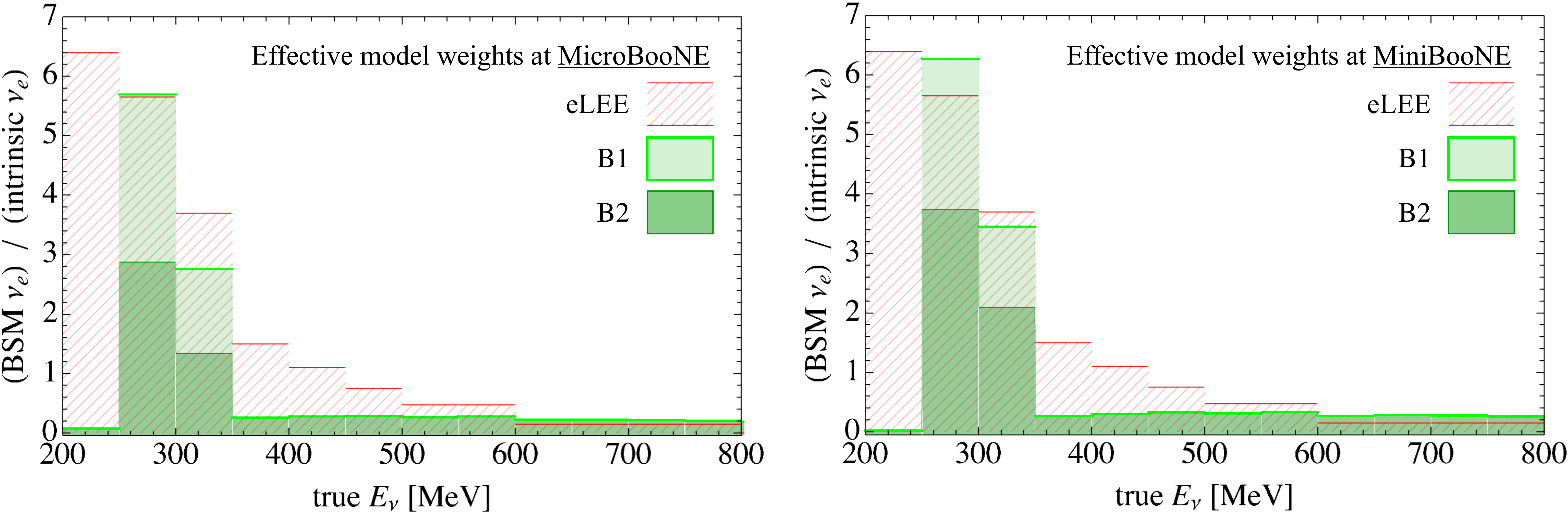}
\caption{The \emph{effective} BSM $\nu_e$ fluxes at the MicroBooNE (left) and MiniBooNE (right) detectors from the BNB normalized by the expected intrinsic $\nu_e$ background flux in the SM, as predicted by three different models: the \emph{ad hoc} empirical LEE model used in the MicroBooNE analyses (red hatched histogram); the quasi-sterile benchmark model B1 defined in~(\ref{B1}) and displayed in figure~\ref{B1MBfit} (light green histogram); and the quasi-sterile benchmark model B2 defined in~(\ref{B2}) and displayed in figure~\ref{B2MBfit} (dark green histogram). For B1 and B2, the `effective' BSM $\nu_e$ fluxes include $\nu_\mu\to\nu_e$ oscillations as well as intrinsic $\nu_e$ disappearance.\label{ModelWeights}}
\end{figure}

While at face value these results appear to strongly disfavor $\nu_e$ appearance interpretations of the MiniBooNE low energy excess, we contend that such a general conclusion is premature. In particular, we highlight several caveats that could invalidate the generality of MicroBooNE's exclusion of $\nu_\mu\to\nu_e$ oscillation hypotheses addressing the MiniBooNE~anomaly:
\begin{enumerate}
\item\label{muBcaveat1}
The observed $\nu_e$ rates in the MicroBooNE BNB data were significantly lower than the background expectation for the three most sensitive channels --- the CCQE Deep-Learning search, for example, found a $p$-value of $p=0.014$ for the background-only hypothesis. This led to stronger than expected exclusions of the empirical LEE model signal strength.  Assuming this discrepancy between data and background expectation was due to a downward statistical fluctuation, it is likely that more data will lead to a weakening of MicroBooNE's constraints. On the other hand, if tensions persist with more data, this would point to a possible mismodelling of expected backgrounds, which would require a revision of the MicroBooNE bounds. Either way, it is unlikely that upcoming MicroBooNE analyses with twice as much beam luminosity will lead to tighter bounds on models predicting BSM $\nu_e$ appearance at MiniBooNE and MicroBooNE.
\item\label{muBcaveat2}
The MicroBooNE analyses tested only the very specific empirical LEE model, which was an \emph{ad hoc} $\nu_e$ excess flux obtained by `reverse-engineering' the MiniBooNE $\nu_e$-like excess. The MicroBooNE collaboration chose to normalize the empirical LEE flux relative to the \emph{intrinsic} $\nu_e$ flux (i.e., the background $\nu_e$ component in the BNB originating from decays of muons and kaons produced at the source). Figure~\ref{ModelWeights} shows the expected empirical LEE $\nu_e$ spectrum at MicroBooNE taken from figure~2 of~\cite{MicroBooNE-ml-2021jwr}; for each energy bin, the empirical LEE flux is weighted relative to the intrinsic $\nu_e$ flux integrated over that same bin. The fact that the reported MicroBooNE exclusions were cast in terms of the empirical LEE signal strength constitutes perhaps the most important caveat concerning these results: \emph{the mapping of the MiniBooNE excess to a truth-level $\nu_e$ spectrum is not injective}, and in fact other $\nu_e$ spectra departing considerably from the empirical LEE model can also provide good fits to the MiniBooNE excess (see also~\cite{Arguelles-ml-2021meu}). We will come back to this point shortly.
\item\label{muBcaveat3}
The MicroBooNE analyses assumed that the empirical LEE $\nu_e$ flux, when normalized to the intrinsic $\nu_e$ flux, was identical at MiniBooNE and MicroBooNE\@. This simplified assumption fails to capture important features of BSM oscillation models. For instance, the fact that MicroBooNE's baseline is about 70\,m shorter than MiniBooNE's  could have a non-negligible effect on $\nu_\mu\to\nu_e$ oscillation probabilities. Furthermore, intrinsic $\nu_e$ disappearance could be non-negligible at MicroBooNE (see also~\cite{Denton-ml-2021czb}), especially if the Gallium anomalies are indeed an indication of large $\nu_e$ disappearance at $L/E_{\nu_e}\sim\mathcal{O}(\text{m}/\text{MeV})$. In this case, MicroBooNE's constraints on a realistic oscillation model would be less severe than those implied from the empirical LEE signal strength due to a reduction of the intrinsic $\nu_e$ backgrounds. (Note that a non-negligible $\nu_e$ disappearance probability would not be as consequential for the MiniBooNE excess because the backgrounds from intrinsic $\nu_e$ CCQE scattering events at MiniBooNE are sub-dominant in the low energy region $E_\nu\big|_\text{\tiny MB}\lesssim 400$\,MeV.)
\end{enumerate}

The triple-resonance model presented in section~\ref{FullModel} provides a clear example of caveats~\ref{muBcaveat2} and~\ref{muBcaveat3} above. Figure~\ref{ModelWeights} (left panel) contrasts the empirical LEE $\nu_e$ flux at MicroBooNE with the \emph{effective} $\nu_e$ flux predicted by the quasi-sterile benchmarks B1 and B2 specified in~(\ref{LSNDparam}) and (\ref{B1}),~(\ref{B2}). In order to provide an `apples-to-apples' comparison with the empirical LEE model, the effective $\nu_e$ fluxes for B1 and B2 shown in figure~\ref{ModelWeights} include not only $\nu_\mu\to\nu_e$ appearance, but also intrinsic $\nu_e$ disappearance (which is predominantly due to $\nu_e$-to-sterile oscillations with an averaged disappearance probability of $\overline{P}_{\nu_e\to S_0} \approx 1/2\times|U_{e 1}|^2 \sin^22\theta_{_{\!S_0}} \approx 11\%$ away from the resonant region). The effects of $\nu_\mu$ disappearance, which would affect the normalization of the $\nu_\mu$ flux and consequently the constrained $\nu_e$ predictions, have been ignored. These effects are small away from the resonance, where they are dominated by $\nu_\mu$-to-sterile oscillations with an averaged $\nu_\mu$ disappearance probability of $\overline{P}_{\nu_\mu\to S_0} \approx 1/2\times|U_{\mu 1}|^2 \sin^22\theta_{_{\!S_0}} \approx 2\%$. Around the resonant region, $\nu_\mu$ disappearance is more significant --- averaging to $\mathcal{O}(10\%)$ for the benchmarks B1 and B2 --- but still within the $\nu_\mu$ flux uncertainty at MicroBooNE\@. A more comprehensive study including the effects of $\nu_\mu$ disappearance are deferred to a future study.

Figure~\ref{ModelWeights} (right panel) contrasts the empirical LEE, B1, and B2 predictions for the BSM $\nu_e$ flux at \underline{Mini}BooNE normalized to the intrinsic $\nu_e$ flux. We can compare these predictions with the ones for \underline{Micro}BooNE shown in the left panel: while the empirical LEE prediction is, by definition, identical at MicroBooNE and MiniBooNE, the predictions from the B1 and B2 quasi-sterile benchmarks are quantitatively different. This is due to the difference in baselines between MicroBooNE and MiniBooNE mentioned in caveat~\ref{muBcaveat3} above. In particular, this difference is more pronounced for B2 because this benchmark does not saturate the appearance probability at the resonance, and MicroBooNE's shorter baseline takes the $\nu_\mu\to\nu_e$ oscillation probability further away from the saturation condition relative to MiniBooNE (see~(\ref{Thetaimax}) and discussion around it).

\begin{figure}
\centering
\includegraphics[width=1.0\textwidth]{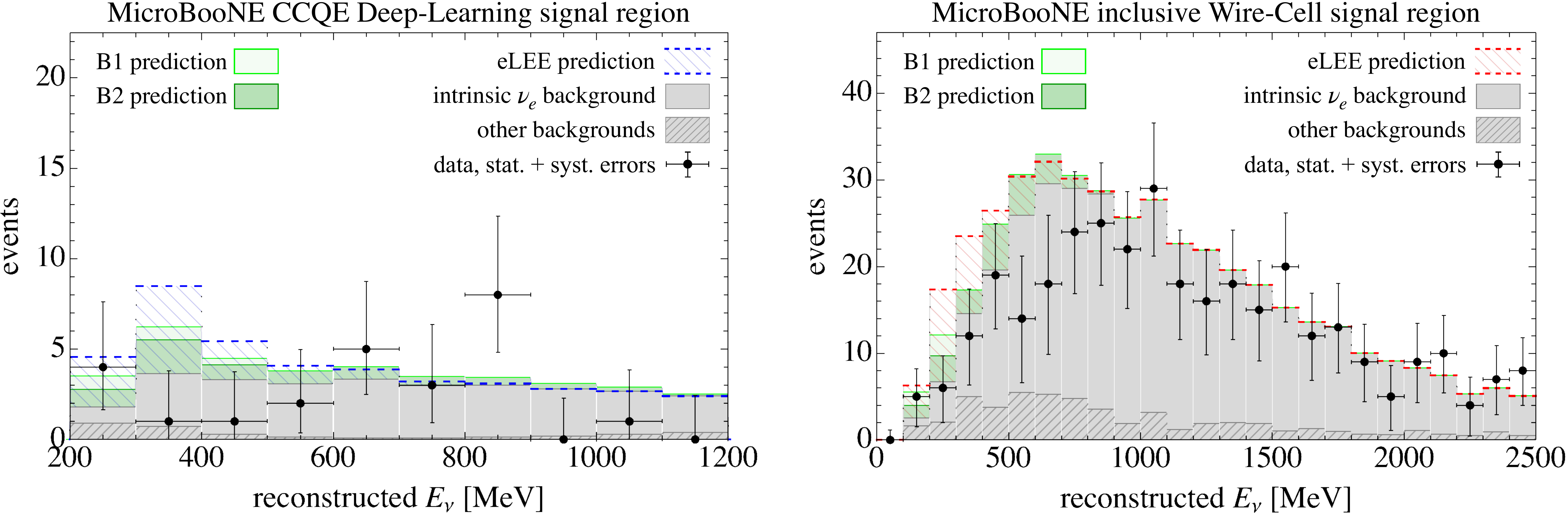}
\caption{Expected $\nu_e$ event rates for two MicroBooNE searches: the CCQE Deep-Learning~\cite{MicroBooNE-ml-2021jwr} (left) and the inclusive Wire-Cell~\cite{MicroBooNE-ml-2021nxr} (right) signal regions. The gray histograms denote the expected backgrounds obtained by the MicroBooNE collaboration in each analysis. In particular, the intrinsic $\nu_e$ background rate shown in gray is the one expected in the SM and does not assume BSM $\nu_e$ oscillations. The hatched histograms (blue on left panel and red on right panel) are the $\nu_e$ event rate prediction from the empirical LEE model used in the MicroBooNE analyses. The light and dark green histograms denote approximate $\nu_e$ event rate predictions from the B1 and B2 quasi-sterile benchmarks (see text for details). Since intrinsic $\nu_e$ disappearance has not been included in the background histograms, this effect has been absorbed in the green histograms for the B1 and B2 predictions. The data error bars include statistical and systematic uncertainties. \label{MicroBooNErates}}
\end{figure}

It is difficult, however, to draw any firm conclusion from the comparisons of truth-level $\nu_e$ rate predictions shown in figure~\ref{ModelWeights}.
Smearing and bias effects in energy reconstruction, and the rapid reduction of event selection efficiency towards lower energies, make it non-trivial to infer which part of the $\nu_e$ signal spectrum dominates the expected $\nu_e$ event rates in the various MicroBooNE analyses. Unlike MiniBooNE's public data releases, the MicroBooNE collaboration has not released sufficient information to allow one to perform an end-to-end statistical analysis of the consistency of an \emph{arbitrary} oscillation model with the MicroBooNE data.

The public data release for the CCQE Deep-Learning analysis~\cite{hepdata.114859.v1}, for instance, provided a Monte Carlo $\nu_e$ event sample~\cite{hepdata.114859.v1-ml-t7} that can be re-weighted to obtain \emph{unconstrained} $\nu_e$ event rate predictions from an arbitrary model. Using these tools, we have obtained the unconstrained $\nu_e$ event rate predicted by the B1 and B2 benchmarks in the CCQE Deep-Learning signal region. However, with the information in the public data release alone, we cannot proceed beyond this point. Specifically, no tools were provided to construct the full covariance matrix for an arbitrary model; without it, one cannot obtain the \emph{constrained} $\nu_e$ event rate prediction, nor its associated $p$-value. Given these limitations, we instead took a heuristic approach: we inferred a bin-by-bin re-weighting procedure that reproduced the constrained $\nu_e$ rates given in~\cite{hepdata.114859.v1-ml-t1}. We applied this same re-weighting procedure to our B1 and B2 benchmarks to obtain their expected constrained $\nu_e$ rates, shown in figure~\ref{MicroBooNErates} (left panel). While we still could not obtain their corresponding $p$-values without a covariance matrix, inspection of figure~\ref{MicroBooNErates} (left) allows us to at least draw a qualitative conclusion from it: the predictions from the quasi-sterile benchmark models are in significantly less tension with MicroBooNE's data than the empirical LEE model's prediction. Indeed, it is conceivable that these benchmarks are not incompatible with data at a high enough confidence level to be excluded.

As for the inclusive Wire-Cell analysis, its public data release to date~\cite{hepdata.114862.v1} provided smearing tables to map true neutrino energy into reconstructed neutrino energy~\cite{hepdata.114862.v1-ml-t5}, as well as event selection efficiencies as a function of true neutrino energy~\cite{hepdata.114862.v1-ml-t3}, which in principle would allow one to obtain the unconstrained $\nu_e$ event rate for an arbitrary model. However, similarly to the public data release for the CCQE Deep-Learning analysis, no tools were provided to allow one to construct the covariance matrix for such an arbitrary model, without which one cannot obtain the constrained $\nu_e$ event rate. Furthermore, the event selection efficiencies provided in the public data release were binned too coarsely in the low energy region to be useful. Therefore, here too we resorted to a semi-heuristic bin re-weighting procedure that reproduced the binned ratio of fully contained, constrained event rate predictions~\cite{hepdata.114862.v1-ml-t1} for the empirical LEE model relative to the intrinsic $\nu_e$ backgrounds. In figure~\ref{MicroBooNErates} (right panel) we show the expected events rates in the inclusive Wire-Cell signal region obtained from applying this semi-heuristic procedure to the B1 and B2 benchmarks. Again, without the means to obtain the covariance matrices for these benchmark models, we could not estimate a $p$-value for the B1 and B2 predictions shown in figure~\ref{MicroBooNErates} (right). However, our prior qualitative conclusion from the CCQE Deep-Learning analysis is not contradicted here: the quasi-sterile oscillation model can provide a good explanation for the MiniBooNE excess while being consistent with the recent constraints from MicroBooNE\@. Further data will hopefully make this situation less ambiguous.

\subsection{Implications for solar neutrinos}\label{SolarNU}

The triple-resonance mechanism that explains the MiniBooNE excess should considerably alter neutrino propagation and flavor conversion in the Sun. Several compounding effects come into play. First, the quasi-sterile resonant energies in the solar core are expected to fall in the range $E^\text{res}_{\nu_i}\big|_{\odot}\sim\mathcal{O}(\rho_\text{\tiny MB}/\rho_{\odot})\,E^\text{res}_{\nu_i}\big|_\text{\tiny MB}\sim\mathcal{O}(\text{MeV})$, which is of the same magnitude as the resonant energy of the standard MSW effect. Second, in the vicinity of the resonant energies $E^\text{res}_{\nu_{1,2}}\big|_{\odot}$, the effective active mass-squared splitting $\delta m_{21 \text{,\,eff\,}}^2$ in the solar medium will deviate by $\mathcal{O}(1-10)$ from the SM expectation, hence altering the MSW effect itself. Third, the quasi-sterile neutrinos $S_i$ ($i=1,2,3$) could have short mean-free-paths in the solar medium due to their vector portal interactions with ordinary matter;  this would cause decoherence of superpositions of active and quasi-sterile neutrino states.

However, precise predictions can only be made once a concrete model is specified for the dark vector mediator responsible for the BSM matter potential. In particular, the quasi-sterile resonant energies in the solar core will depend not only on the ratio of dirt-to-solar core densities, $\rho_\text{\tiny MB}/\rho_{\odot}$, but also on the specific couplings of the dark mediator to electrons, neutrons, and protons, since the relative neutron-to-proton and nucleon-to-electron compositions in the solar core differ by $\mathcal{O}(1)$ from the shallow Earth crust composition through which neutrinos propagate in accelerator experiments. In addition, in order to compute the quasi-sterile neutrinos' mean-free-path in the solar medium, the mediator's mass and couplings are needed. Finally, the model-dependent solar emission of quasi-sterile neutrinos could lead to scattering signals in neutrino and dark matter experiments such as SK, SNO, Borexino, and XENON1T (see also~\cite{Harnik-ml-2012ni,Cerdeno-ml-2016sfi,Gann-ml-2021ndb,Goldhagen-ml-2021kxe}).

Our preliminary investigations indicate that is it possible to build a dark sector UV completion that is experimentally viable and leads to predictions compatible with existing solar neutrino data. This is however beyond the scope of this work; a complete study of implications of this model to active and quasi-sterile solar neutrinos will appear in a follow-up paper~\cite{future}.

\section{Summary and outlook}
\label{Conclusion}

An inevitable consequence of dark sector models with quasi-sterile neutrinos are new matter effects in neutrino propagation through a medium, which can lead to two generic effects. The first is resonant active-to-sterile neutrino oscillations around a resonant energy $E^\text{res}_\nu$ determined by the medium's forward scattering potential. This effect occurs for either neutrinos or antineutrinos, but not both, and the choice is determined by the sign of the effective matter potential. The second effect, which affects both neutrinos and antineutrinos, is an energy- and medium-dependent modification of the active neutrino mass spectrum. With suitable parameter choices, this modification may disappear in the asymptotic limits $E_\nu\ll E^\text{res}_{\nu}$ and $E_\nu\gg E^\text{res}_\nu$, but it is inevitable and increasingly pronounced as $E_\nu$ approaches the resonant energy. Generic realizations of these effects are obviously not phenomenologically viable given the swath of experimental measurements targeting different neutrino flavors, energies, and~baselines.

In this paper, we have studied a specific realization of a model with quasi-sterile neutrinos and resonant oscillations that can address the short-baseline $\nu_\mu\to\nu_e$ anomalies while preserving the active neutrino mass spectrum in regimes probed by long-baseline experiments. We have also argued that the recent $\nu_e$ measurements by MicroBooNE likely do not exclude this model. Additional data from MicroBooNE expected in the near future is unlikely to unambiguously resolve the initial tensions between prediction and observations; however, in the longer term, the SBN program, including MicroBooNE, SBND, and ICARUS, will be able to fully test our proposed explanation of the MiniBooNE and LSND anomalies.

Further, while this model can accommodate non-negligible $\nunubar_{\!e}$ disappearance probabilities, it does not address the significant tensions between the existing $\nunubar_{\!e}$ disappearance measurements. Proposed experiments such as IsoDAR@Yemilab~\cite{Alonso-ml-2021jxx,Alonso-ml-2021kyu} will be rid of large systematic flux uncertainties, and will be able to reconstruct the antineutrinos energy and propagated distance with high precision on an event-by-event basis. As such, it should be able to resolve the ambiguities and inconsistencies that currently plague the $\nunubar_{\!e}$ disappearance experimental landscape. Being an \emph{antineutrino} experiment, IsoDAR@Yemilab will not be able to probe the resonant mechanism within the model proposed in this study; nonetheless, it will be able to test this model through its non-resonant $\bar\nu_e$ oscillation~predictions.

The resonant mechanism of this quasi-sterile neutrino model should also have significant consequences for solar neutrinos, and offer testable predictions for active solar neutrino emission, as well as quasi-sterile solar neutrino emission which could lead to distinct signals in dark matter and solar neutrino experiments. A comprehensive study of the BSM solar neutrino phenomenology in this context is deferred to a future publication~\cite{future}.

\acknowledgments

DSMA benefited from conversations with the late Ann Nelson on model building aspects of this study; Ann is sorely missed. DSMA also acknowledges the Aspen Center for Physics (ACP) where this work was partially developed --- the ACP is supported by National Science Foundation grant PHY-1607611. We are grateful to En-Chuan Huang for useful explanations regarding the MiniBooNE data, and Janet Conrad and Mark Ross-Lonergan for discussions regarding the recent MicroBooNE analyses.
This work was supported by the Los Alamos National Laboratory LDRD program, and by the DOE Office of Science High Energy Physics under contract number DE-AC52-06NA25396.





\end{document}